\documentclass[a4paper,11pt]{article}
\usepackage{jheppub}
\usepackage{hyperref}       
\hypersetup{
    colorlinks=true,
    linkcolor=blue,
    filecolor=magenta,      
    urlcolor=blue,
}
\usepackage{url}
\usepackage{booktabs}       
\usepackage{amsmath}       
\usepackage{nicefrac} 
\usepackage{microtype} 
\usepackage{lipsum}
\usepackage{graphicx}
\usepackage{amsmath}
\usepackage{multirow}
\usepackage{xspace}
\usepackage{setspace}
\usepackage{glossaries-prefix}
\usepackage[capitalise,noabbrev]{cleveref}
\usepackage{todonotes}
\usepackage{float}
\usepackage[toc,page]{appendix}
\usepackage{tcolorbox}
\usepackage{aas_macros}

\renewcommand{\vec}[1]{\boldsymbol{#1}}

\begin{document}

\title{Applications of Deep Learning to physics workflows}
\onecolumn


\author[1]{Manan Agarwal,}
\author[2]{Jay Alameda,}
\author[3,1]{Jeroen Audenaert,}
\author[4]{Will Benoit,}
\author[5]{Damon Beveridge,}
\author[6]{Meghna Bhattacharya,}
\author[5]{Chayan Chatterjee,}
\author[1]{Deep Chatterjee,}
\author[7]{Andy Chen,}
\author[4]{Muhammed Saleem Cholayil,}
\author[7]{Chia-Jui Chou,}
\author[5]{Sunil Choudhary,}
\author[4]{Michael Coughlin,}
\author[8]{Maximilian Dax,}
\author[]{Aman Desai,}
\author[9]{Andrea Di Luca,}
\author[10]{Javier Mauricio Duarte,}
\author[11]{Steven Farrell,}
\author[6]{Yongbin Feng,}
\author[12]{Pooyan Goodarzi,}
\author[1]{Ekaterina Govorkova,}
\author[13]{Matthew Graham,}
\author[10]{Jonathan Guiang,}
\author[1]{Alec Gunny,}
\author[5]{Weichangfeng Guo,}
\author[14]{Janina Hakenmueller,}
\author[6]{Ben Hawks,}
\author[15]{Shih-Chieh Hsu,}
\author[16]{Pratik Jawahar,}
\author[11]{Xiangyang Ju,}
\author[1]{Erik Katsavounidis,}
\author[1]{Manolis Kellis,}
\author[15]{Elham E Khoda,}
\author[17]{Fatima Zahra Lahbabi,}
\author[14]{Van Tha Bik Lian,}
\author[18]{Mia Liu,}
\author[2]{Konstantin Malanchev,}
\author[1]{Ethan Marx,}
\author[1]{William Patrick McCormack,}
\author[5]{Alistair McLeod,}
\author[1]{Geoffrey Mo,}
\author[1]{Eric Anton Moreno,}
\author[1]{Daniel Muthukrishna,}
\author[2]{Gautham Narayan,}
\author[11]{Andrew Naylor,}
\author[19]{Mark Neubauer,}
\author[20]{Michael Norman,}
\author[4]{Rafia Omer,}
\author[6]{Kevin Pedro,}
\author[21]{Joshua Peterson,}
\author[22]{Michael P{\"u}rrer,}
\author[23]{Ryan Raikman,}
\author[24]{Shivam Raj,}
\author[1]{George Ricker,}
\author[]{Jared Robbins,}
\author[25]{Batool Safarzadeh Samani,}
\author[14]{Kate Scholberg,}
\author[15]{Alex Schuy,}
\author[20]{Vasileios Skliris,}
\author[1]{Siddharth Soni,}
\author[13]{Niharika Sravan,}
\author[20]{Patrick Sutton,}
\author[26]{Victoria Ashley Villar,}
\author[2]{Xiwei Wang,}
\author[5]{Linqing Wen,}
\author[27]{Frank Wuerthwein,}
\author[6]{Tingjun Yang,}
\author[28]{Shu-Wei Yeh}
\affiliation[1]{Massachusetts Inst. of Technology (US)}
\affiliation[2]{University of Illinois at Urbana-Champaign (US)}
\affiliation[3]{KU Leuven (BE)}
\affiliation[4]{University of Minnesota (US)}
\affiliation[5]{The University of Western Australia (AU)}
\affiliation[6]{Fermi National Accelerator Lab. (US)}
\affiliation[7]{National Yang Ming Chiao Tung University (TW)}
\affiliation[8]{Max Planck Institute for Intelligent Systems (DE)}
\affiliation[9]{Universita degli Studi di Trento and INFN (IT)}
\affiliation[10]{Univ. of California San Diego (US)}
\affiliation[11]{Lawrence Berkeley National Lab. (US)}
\affiliation[12]{Univ. of California, Riverside (US)}
\affiliation[13]{California Inst. of Technology (US)}
\affiliation[14]{Duke University (US)}
\affiliation[15]{University of Washington Seattle (US)}
\affiliation[16]{University of Manchester (GB)}
\affiliation[17]{Universite Hassan II, Ain Chock (MA)}
\affiliation[18]{Purdue University (US)}
\affiliation[19]{University of Illinois at Urbana Champaign (US)}
\affiliation[20]{Cardiff University (GB)}
\affiliation[21]{University of Wisconsin - Madison (US)}
\affiliation[22]{University of Rhode Island (US)}
\affiliation[23]{Carnegie Mellon University (US)}
\affiliation[24]{Catholic University of America (US)}
\affiliation[25]{University of Sussex (GB)}
\affiliation[26]{Pennsylvania State University (US)}
\affiliation[27]{San Diego Supercomputer Center (US)}
\affiliation[28]{National Tsing Hua University (TW)}

\maketitle

\clearpage
\begin{abstract}
\section{Modern large-scale physics experiments create datasets with sizes and streaming rates that can exceed those from industry leaders such as Google Cloud and Netflix.
Fully processing these datasets requires both sufficient compute power and efficient workflows.
Recent advances in Machine Learning (ML) and Artificial Intelligence (AI) can either improve or replace existing domain-specific algorithms to increase workflow efficiency.
Not only can these algorithms \textit{improve} the physics performance of current algorithms, but they can often be executed more quickly, especially when run on coprocessors such as GPUs or FPGAs.
In the winter of 2023, MIT hosted the Accelerating Physics with ML at MIT workshop, which brought together researchers from gravitational-wave physics, multi-messenger astrophysics, and particle physics to discuss and share current efforts to integrate ML tools into their workflows.
The following white paper highlights examples of algorithms and computing frameworks discussed during this workshop and summarizes the expected computing needs for the immediate future of the involved fields.}
\tiny
\end{abstract}


\clearpage

\section{Introduction}


Machine learning (ML) and artificial intelligence (AI) is a rapidly developing field that has given rise to physics-relevant techniques such as classification, tagging, noise reduction, event reconstruction, and anomaly detection.
As workflows in experimental physics become increasingly saturated by ML, it is important to maximize computational efficiency to reduce both processing latency and computing demands.
One way to increase the efficiency of ML algorithms is to use heterogeneous computing frameworks that incorporate coprocessor hardware such as GPUs and FPGAs.


While large-scale computing facilities in the US have provisioned modern hardware dedicated for
scientific analysis, there is a lack of standardization of tools to efficiently use these heterogeneous
resources. High performance computing centers (HPC centers), such as those present at the National Energy Research Scientific Computing Center (NERSC) or the San Diego Supercomputer Center (SDSC), have large GPU allocations capable of very significant compute. Despite that, much of the computing infrastructure has been focused on the deployment of large-scale simulations and calculations in domains including Lattice QCD and astrophysical modeling. While HPC centers have enabled enormous advancements in those domains, there has been little use of these systems for real-time operations of big physics experiments. Recent advancements in the use of AI within physics workflows have demonstrated enormous speedups and improved algorithm performance. As a result, there is a growing interest in utilizing large-scale heterogeneous computing resources where substantial computational speedups are possible. A potential synergistic opportunity has emerged where the large-scale deployment of physics workflows on heterogeneous HPC systems can substantially enhance the computational abilities of next-generation physics experiments, leading to a wealth of possibilities. 

There are, however, some hurdles to the use of HPC centers for real-time physics experiments. The dynamic balancing of CPU to GPU resources, the deployment of different algorithms to different GPUs, and the use of industry tools to control large-scale computation have had limited use in HPC centers. But with a few adjustments in the design and use of current and next-generation HPC centers, there is a large potential to harness these HPC centers for the large-scale deployment of AI-enhanced real-time and data processing workflows for physics experiments.
To increase community awareness of these ML/AI and computational tools, workshops, such as ``Accelerating Physics with ML at MIT'', and institutions, such as the Institute for Accelerated AI Algorithms for Data Driven Discovery (A3D3), have brought together researchers from different fields to share experiences with various algorithms and computing frameworks.
In this white paper, we highlight the many algorithms that are being developed and their impact in the domains of electromagnetic (EM) astronomy, gravitational wave (GW) astronomy, and high energy physics (HEP). We then discuss the computational demands of these algorithms and build a path towards the computational resources that will enable the large-scale adoption of HPC centers for large physics experiments. 

\pagebreak

\subsection{EM Summary}
\label{sec:em_sec}
\begin{figure}[h]
    \centering
    \includegraphics[width=0.45\textwidth]{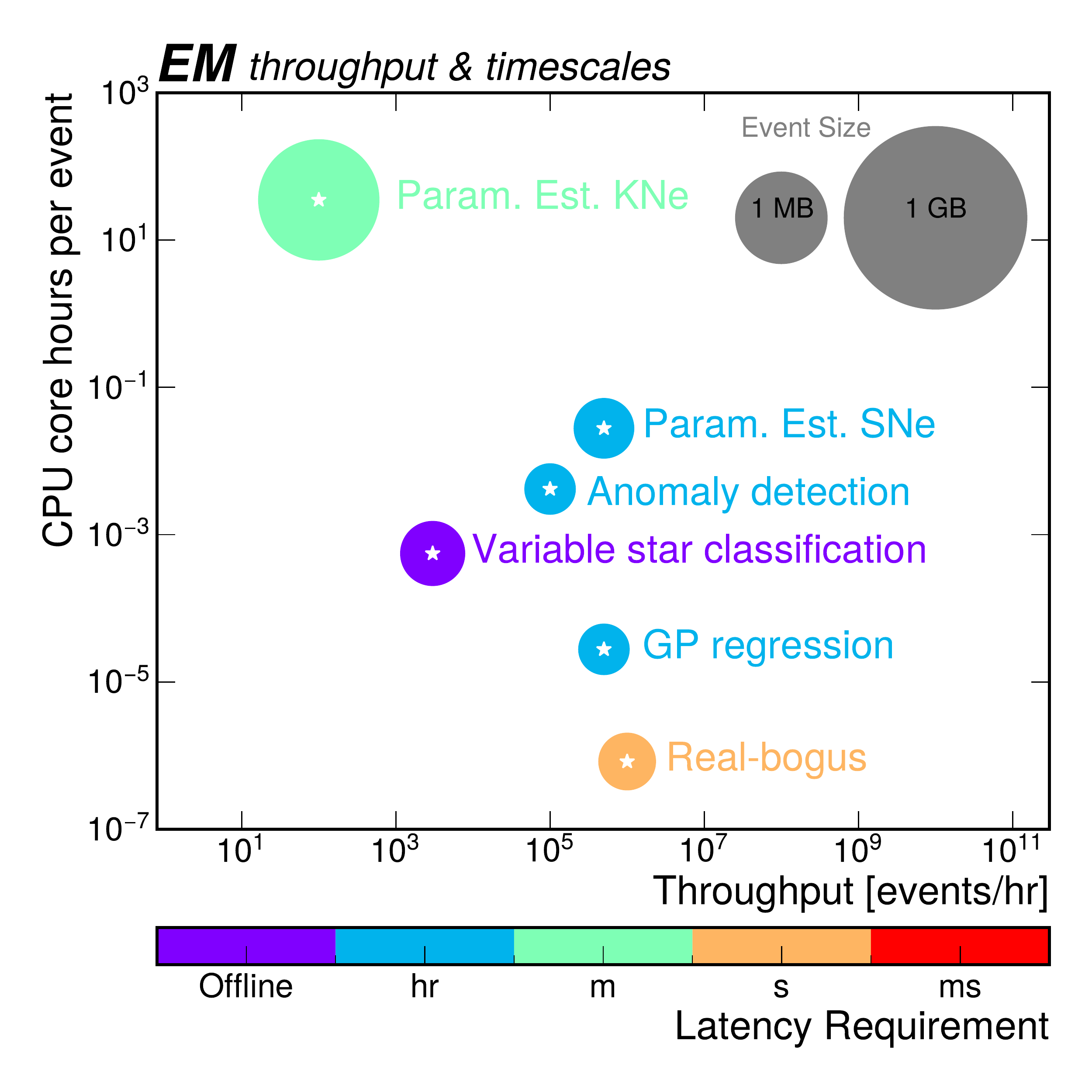}
    \includegraphics[width=0.45\textwidth]{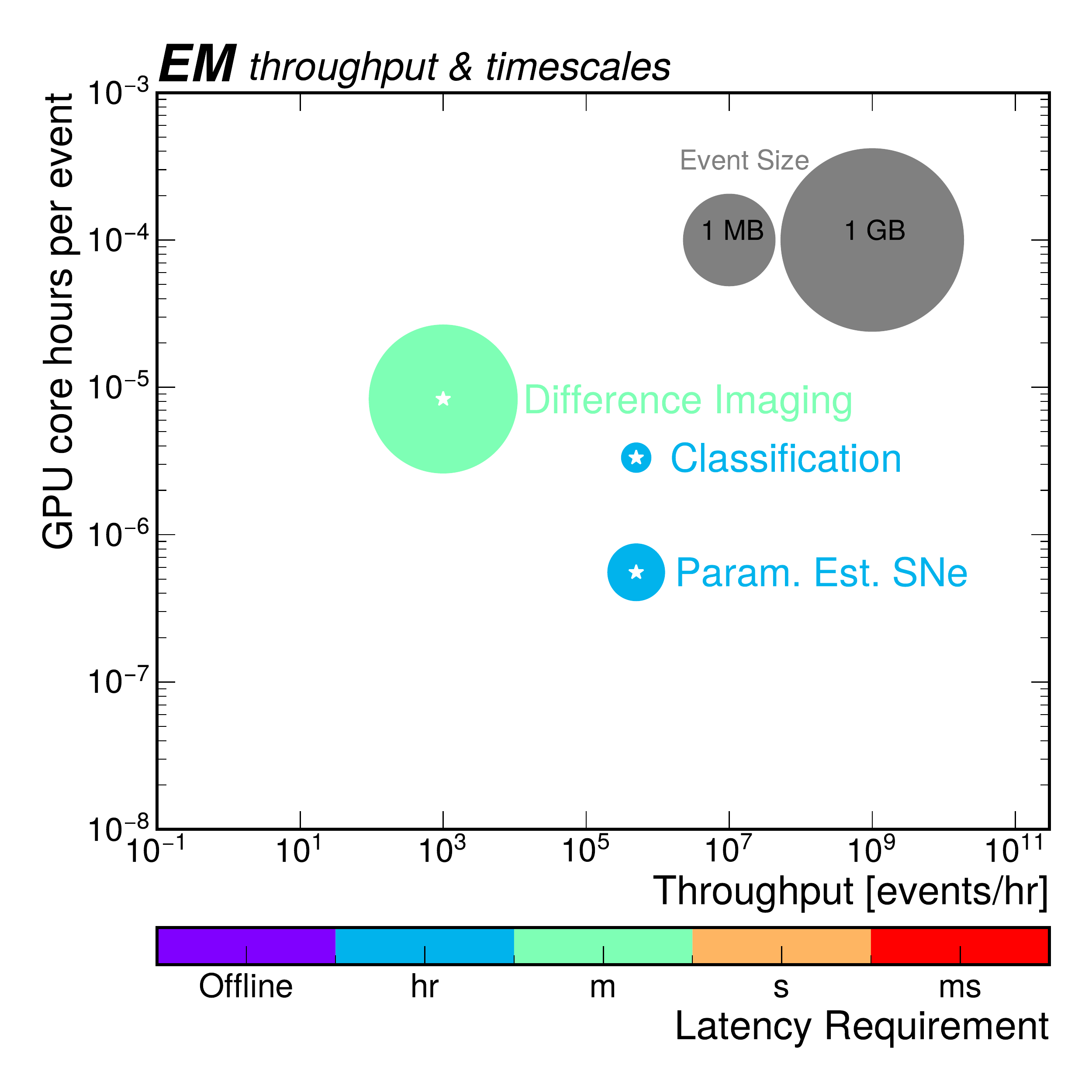}
    \caption{Throughput and CPU (left) and GPU (right) core hours per event for highlighted EM workflows. The size of the circles represents typical event sizes, and their colors represent latency requirements (per event) for the workflows. The computational requirements were estimated using the following sources: parameter estimation for kilonovae \citep{Pang2022KNe}, parameter estimation for supernovae \citep{Boone2021Parsnip,Villar2022SBI}, anomaly detection \citep{Muthukrishna2022}, variable star classification \citep{Audenaert2021}, GP regression \citep{Boone2019Avocado}, classification \citep{Muthukrishna19RAPID}, real-bogus detection \citep{Duev2019Real-bogus}, difference imaging \citep{Corbett2022DIA}.}
    \label{fig:em_times_and_compute}
\end{figure}

Time-domain astronomy is entering a data revolution as new electromagnetic optical observatories begin to observe more data than ever before. Upcoming large-scale optical surveys such as the Vera C. Rubin Observatory's Legacy Survey of Space and Time (LSST), scheduled to begin observations in 2025, will observe transient alerts at a rate more than an order of magnitude larger than any previous survey \citep{Ivezic2009LSST:Products}. Ongoing surveys, such as the Zwicky Transient Facility (ZTF) \citep{Bellm2019ZTF}, Transiting Exoplanet Survey Satellite (TESS) \citep{TESS_Ricker_2015}, and the Panoramic Survey Telescope and Rapid Response System (PanSTARRS) \citep{Chambers2016Panstarrs}, are already recording millions of transient alerts. To deal with these current data streams, and in preparation for the much larger data stream from LSST, a range of machine learning algorithms are being developed to process, classify, and characterize the transients in these alert streams. 

LSST is expected to record 20 TB of images per night, corresponding to over ten million transient alerts each night. These alert packets will be made available to the community with a latency of 60 seconds. To manage these data volumes, seven \textit{Alert Brokers} are being actively developed (\texttt{ALeRCE} \citep{ALeRCE_Broker}, \texttt{AMPEL} \citep{AMPEL_Broker}, \texttt{ANTARES} \citep{ANTARES_Broker}, \texttt{BABAMUL}, \texttt{Fink} \citep{Fink_Broker}, \texttt{Lasair} \citep{LASAIR_Broker}, \texttt{Pitt-Google}). These brokers are responsible for processing the alert streams from multiple surveys, building a data lake, and providing science-ready access to data for the scientific community. While the brokers will have some machine learning capabilities, they have no requirement for any computational backend, and are not necessarily capable of dealing with large computational algorithms. Currently, there is no standardized platform for computing resources.

For many transient phenomena, it is critical that follow-up observations happen quickly to improve understanding of an object's physical mechanisms. Obtaining detailed follow-up, such as spectroscopy and multi-wavelength photometry shortly after a transient's explosion, provides insights into the progenitor systems and central engine that powers the events. Events such as the shock breakout of a supernova occur on a timescale of seconds to hours, while kilonovae require follow-up observations at timescales less than a day, and longer-lived transients require follow-up within less than a week to understand the physics behind the event. 

In Fig.~\ref{fig:em_times_and_compute}, we plot the computational requirements for some key algorithms in time-domain optical astronomy that may benefit from real-time use of HPC facilities. We discuss some of these key algorithms in the following paragraphs. They follow a typical chain of Alert preparation to identify transients, followed by classification of the transients, and concluding with parameter estimation of the identified transients. 

\textbf{Alert Preparation}: Processing the images from survey telescopes to discover transient sources requires \textit{difference imaging} analysis. These computationally-intensive algorithms have been sped up using GPU acceleration \citep{Corbett2022DIA} and require HPC centers to handle the terabytes of images being observed each night. \textit{Real-bogus} classification algorithms are then run to identify which of the detected transients are real and which are artifacts of instrument noise or other non-astrophysical phenomena \citep{Duev2019Real-bogus}.

\textbf{Classification:} A range of machine learning algorithms are currently being used to classify the different types of alerts coming from real-time data streams. In particular, neural network architectures such as Recurrent Neural Networks and Transformers have shown promise for the \textit{classification} and \textit{anomaly detection} of transients (e.g. \citep{Muthukrishna19RAPID, Muthukrishna2022, Villar2020SuperRAENN, Villar2021_Anomalydetection, SupernnoovaMoller2019, Pimentel2023, Tarek2023}). These algorithms can be run in real time on GPUs. However, many of these algorithms first perform \textit{Gaussian Process Regression} on CPUs for interpolation or data augmentation of the time series before classification \citep{Boone2019Avocado}. Many transient phenomena need to be identified within minutes to days so that follow-up observations can be made with other telescopes while the transient variability is still active. Conversely, variable stars and exoplanets are often periodic and thus do not have the same time-sensitivity for follow-up. \textit{Variable star classification} typically requires computationally-intensive feature extraction processes run on parallel CPUs before running machine learning algorithms (e.g. \citep{Audenaert2021}). The expected CPU and GPU computational needs for the classification and anomaly detection of transients and variable stars are plotted in Fig.~\ref{fig:em_times_and_compute}. 

\textbf{Parameter Estimation:} Once a candidate transient has been identified by a machine learning classifier, real-time parameter estimation can help to identify key physical parameters that enable scientists to make decisions in real time about which events to follow up. Parameter estimations of supernovae (SNe) typically involve costly MCMC analyses (e.g. \citep{Mandel2022}); however, recent approaches use machine learning algorithms such as normalizing flows and neural network autoencoders to significantly speed up the inference of physical parameters (e.g. \citep{Boone2021Parsnip,Villar2022SBI}). Kilonovae (KNe) are extremely rare phenomena, and estimating their parameters currently uses a combination of optical, gravitational wave, and gamma-ray datasets. The combined modeling of these datasets is very computationally expensive \citep{Pang2022KNe} and can thus only be run on a subset of candidate events when a kilonova alert occurs.










\subsection{GW Summary}
\label{sec:gw_sec}
\begin{figure}[h]
    \centering
    \includegraphics[width=0.48\textwidth]{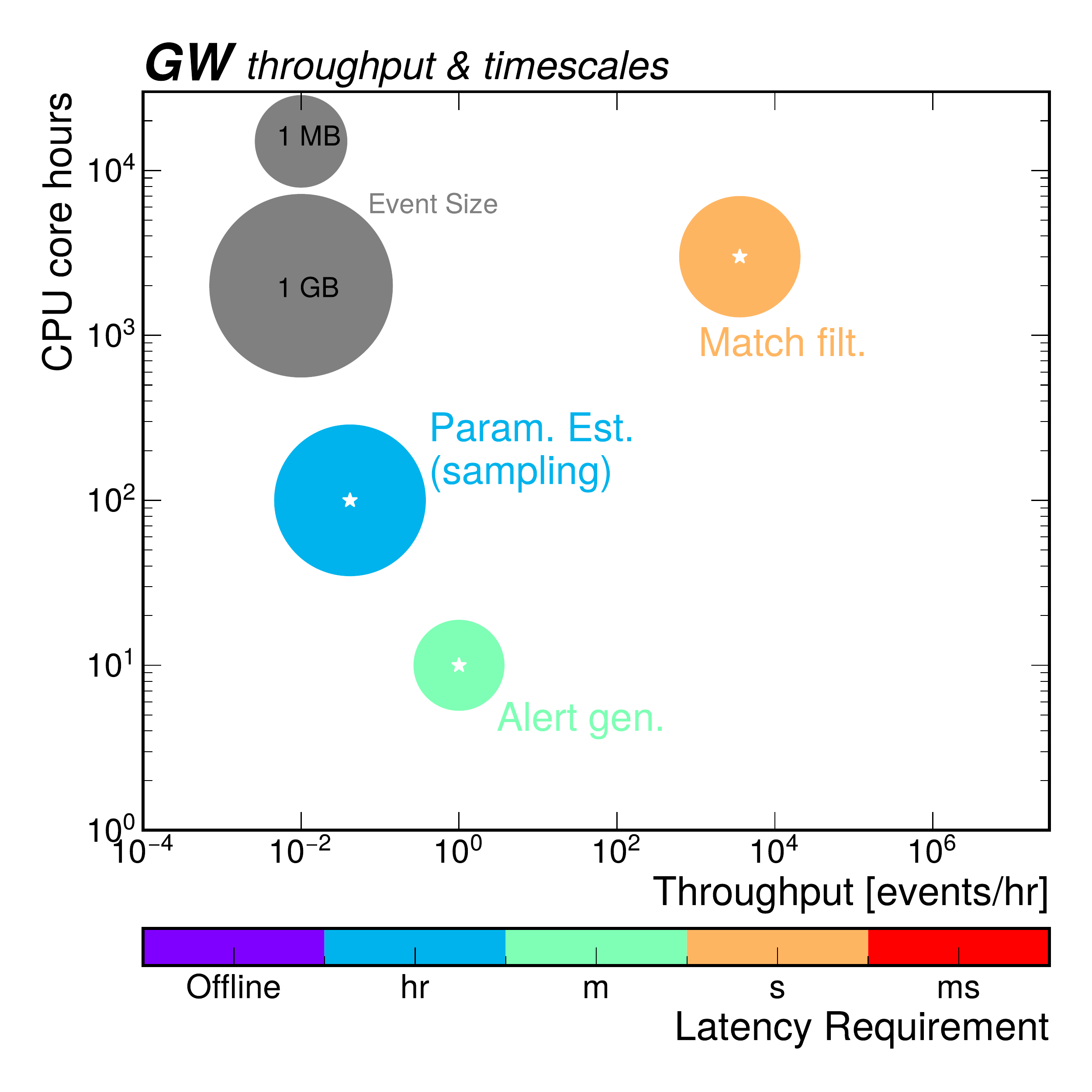}
    \includegraphics[width=0.48\textwidth]{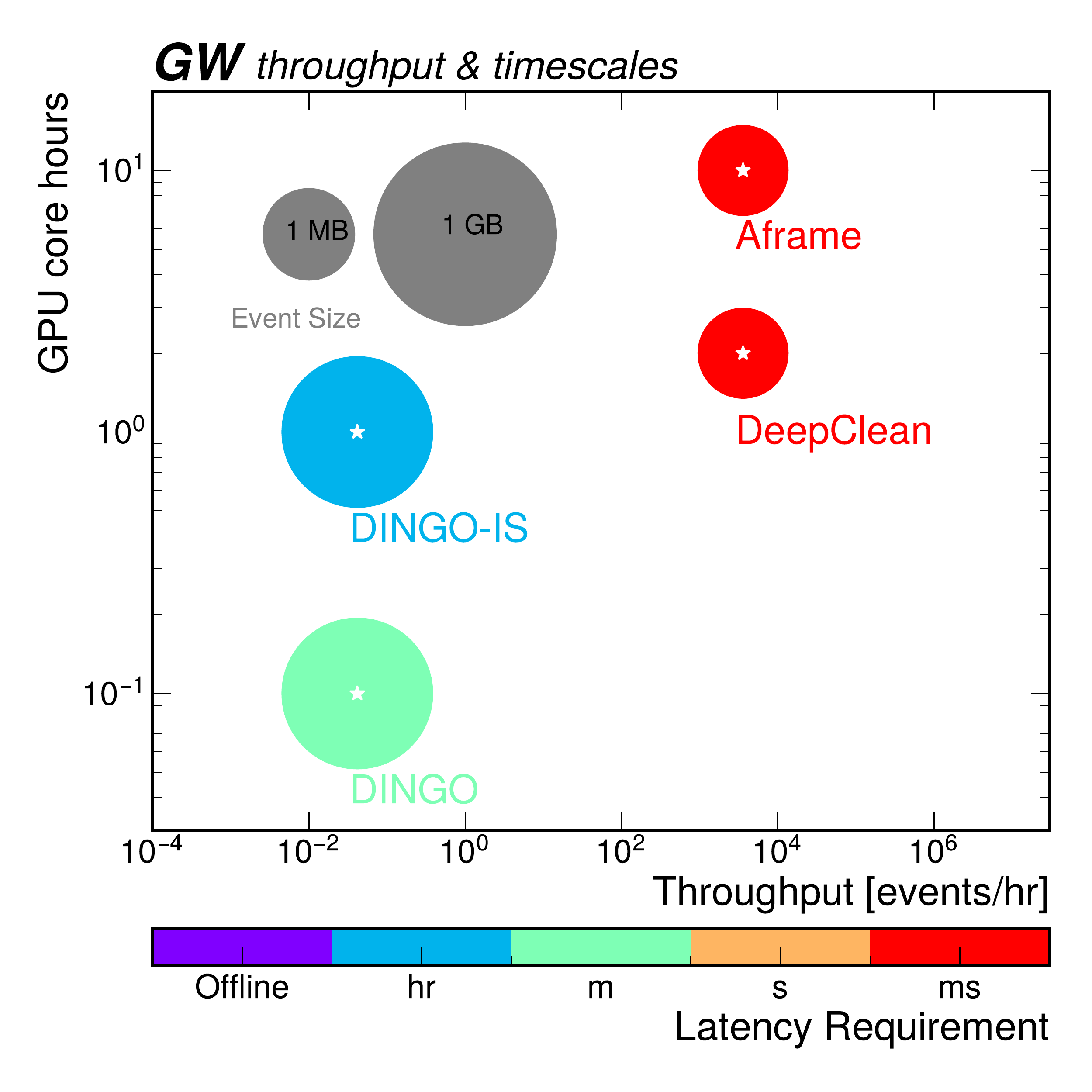}
    \caption{Timescale, compute, and throughput involved in GW low-latency science. Note that for
    online purposes, gravitational-wave data is streamed in chunks of $\sim 1$ second, which
    sets throughput for most searches. Given current astrophysical rates for compact binaries,
    high-significance triggers are expected $\sim 1$ day, which sets the throughout of parameter
    estimation algorithms. A significant discovery is reported and updated in $\approx 3-4$ alerts, which comprise of annotations that help in the EM followup of the candidates.}
    \label{fig:gw_times_and_compute}
\end{figure}

The direct observations of gravitational waves (GWs) in 2015~\cite{Abbott_2016} was a
landmark in physics, leading to the 2017 Nobel Prize and marking another triumph of general
relativity (GR)~\cite{1916AnP...354..769E}. The field has made significant progress since then,
with the number of GW events increasing from 3 to 90 over the last three observing runs~\cite{LIGOScientific:2021djp}. 
The trend is expected to continue in the next LIGO-Virgo-KAGRA observing run which started in May 2023.\footnote{\small{\url{https://observing.docs.ligo.org/plan/}; \\ \url{https://emfollow.docs.ligo.org/userguide/capabilities.html}}}
Combined with the increased GW discoveries, the scope of multi-messenger astronomy is one of the most
interesting and simultaneously challenging topics in astronomy today, as highlighted in the \emph{New messenger, New physics}
theme of the Astro2020 Decadal survey~\cite{astro2020}. The unprecedented increase in discovery rate poses a challenge in terms of
algorithms and compute necessary to scale for future observing runs and next-generation facilities.
In addition, the increasing number of ``interesting'' candidates is overwhelming the joint searches for exotic objects like
kilonovae; there has been no confirmed success since the first binary neutron star, GW170817~\cite{PhysRevLett.119.161101}. 

Looking at the GW landscape over the last five years, it is clear that new algorithms are
needed to keep up with discovering novel signatures in the increasing data volume. The compute
requirements for searching, classifying, and cataloging GW events in the third observing run
(O3) era was already ${\sim}\;0.5$ billion CPU core-hours. The upcoming fourth observing run is likely to
find more sources than the cumulative total discovered until now. Additionally, the sources being discovered
are beyond the ``garden variety'', requiring more accurate, and hence more expensive models. It is becoming
progressively difficult to meet the requirements for the next-generation instruments without a paradigm
shift in algorithms. Machine learning brings promise in various aspects, from noise removal to discovering unmodeled physics. Several avenues,
starting from data cleaning to searches and parameter inference, were discussed in the workshop.
In Fig.~\ref{fig:gw_times_and_compute}, a ballpark estimate of the throughput vs. compute resource
required for core aspects of GW data analysis is shown. The analyses are divided based on their CPU or GPU
resource requirements, comparing
established workflows to ML-based analyses. The latter predominantly use
GPU resources or hybrid architectures with GPUs handling the compute-intensive portion.
Analyses that
take advantage of coprocessors like GPUs can achieve orders of magnitude improvement in terms of inference latency. This will be necessary for the next
generation of ground-based instruments like Cosmic Explorer \cite{evans2021horizon}, and eventually LISA \cite{amaroseoane2017laser}. 
However, benchmarking the results and assessing robustness, especially in an online setting,
and having the infrastructure to efficiently use coprocessors to accelerate analyses is
necessary for adoption into routine real-time GW data analysis. Below is an overview of the broad
areas where ML algorithms have shown promise:

\textbf{Noise subtraction}: Environmental effects couple to the GW detector response in non-trivial
ways. An example is the non-linear coupling of the 60 Hz power line, which results in secondary
bands around the 60 Hz line that are difficult to remove via linear subtraction. However,
the DeepClean algorithm \cite{Ormiston_2020}, a variational autoencoder, has been demonstrated to remove the non-linear
couplings effectively, resulting in an increased range, especially for stellar mass binaries,
without any negative impact on the parameter estimation. The cost of training the network is $\mathcal{O}$(hours)
on a single GPU, and inference is $\mathcal{O}$(ms).

\textbf{Searches}: Matched filtering is the established technique to discover GWs, relying on
$\mathcal{O}$(million) templates and compute resources ranging from several hundred to
thousands of CPU cores. In this domain, the SPIIR team has shown that the use of temporal networks
(CNN + LSTM) can lead to better detection statistics~\cite{beveridge} and can be used for waveform
extraction from detector data~\cite{Chatterjee_2021}.
Similarly,
construction of low-latency
data products such as skymaps using normalizing flows has also been demonstrated \cite{chatterjee2022premerger, chatterjee2022rapid}.
The Aframe project\footnote{\url{https://github.com/ML4GW/aframe}} takes a different approach, using a ResNet
architecture to directly construct a streaming detection statistic starting from the strain data.
The presence of detector glitches is known to cause false alarms in the search for
compact binaries. The training scheme of Aframe employs real detector noise with glitch
injections as well as signals. Inference-as-a-service (IaaS) enables efficient use
of hardware during inference. Regarding unmodeled searches, the MLy search, trained on
white noise bursts, has been shown to recover signals of different morphologies. Preliminary
adoption of IaaS has been carried out for validation and production purposes. These algorithms typically take $\sim$ hours to train on $\mathcal{O}(1)$ GPU(s). Streaming inference, depending
on the rate, may require $\mathcal{O}(1-10)$ GPU(s).

Other problems, such as anomaly detection, take an entirely different approach by
considering unmodeled signals as anomalies. Preliminary work has demonstrated that core-collapse
supernova signals can be discriminated effectively from other known signals and glitch morphologies
in a lower-dimensional embedding. Likewise, distinguishing black hole captures from high-mass, short-lived signals has also been shown to work using variational autoencoders \cite{Guo_2022}.

\textbf{Parameter Estimation}: Amortized simulation-based inference has been successfully demonstrated
in several areas of physics, such as cosmology and high-energy physics \cite{Alsing_2019, 2021AJ....161..262Z,Cranmer_2020}.
The DINGO algorithm \cite{PhysRevLett.127.241103,Dax:2021myb,PhysRevLett.130.171403} performs amortized neural posterior
estimation of binary parameters from observed GW events. DINGO uses normalizing flows to estimate the posterior distribution
at similar accuracy as stochastic sampling techniques. Moreover, DINGO combined with importance sampling \cite{PhysRevLett.130.171403}
(assuming a GW likelihood) corrects for potential neural network inaccuracies, outputs the sample efficiency to directly assess
the robustness of results, and provides an unbiased estimate of the Bayesian evidence.
Given the growing number of discoveries, amortized simulation-based inference offers a pathway toward avoiding the increasing
compute costs associated with stochastic sampling. Training a DINGO BBH network for existing methods takes $\sim 200$
GPU hours. Improvements strategies in training is proposed in Ref. \cite{dax2023flow}. However, inference can be performed within a few minutes. The cost of optional importance sampling is 
$\sim 10$ hours on $\mathcal{O}(100)$ CPU cores, depending on the complexity of the GW waveform model.
Other applications involving low-latency inference on mass parameters directly from time-domain data have been
shown to work with an autoencoder network \cite{mcleod2022rapid}.

\subsection{HEP Summary}
\label{sec:hep_sec}
\begin{figure}[h]
    \centering
    \includegraphics[width=0.45\textwidth]{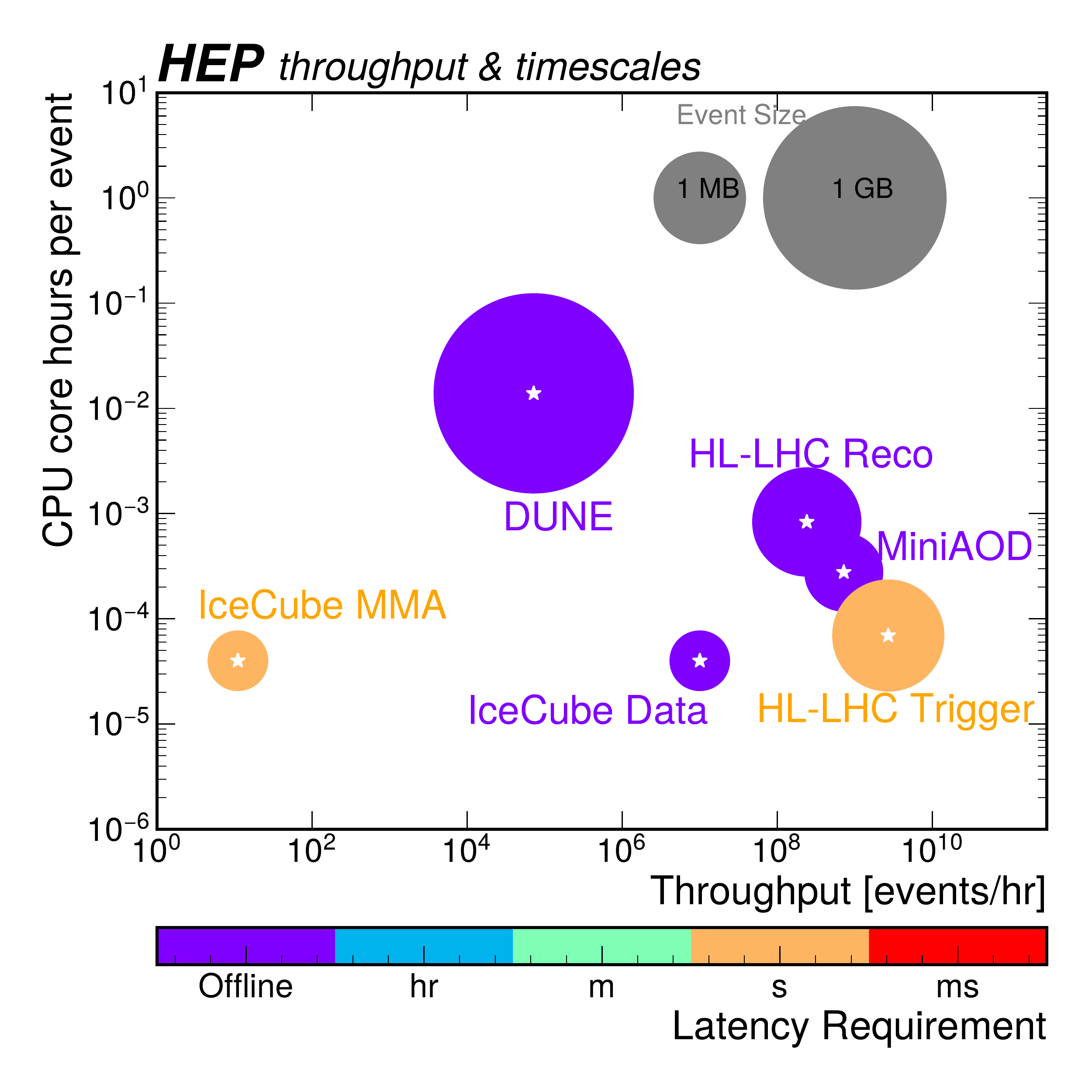}
    \includegraphics[width=0.45\textwidth]{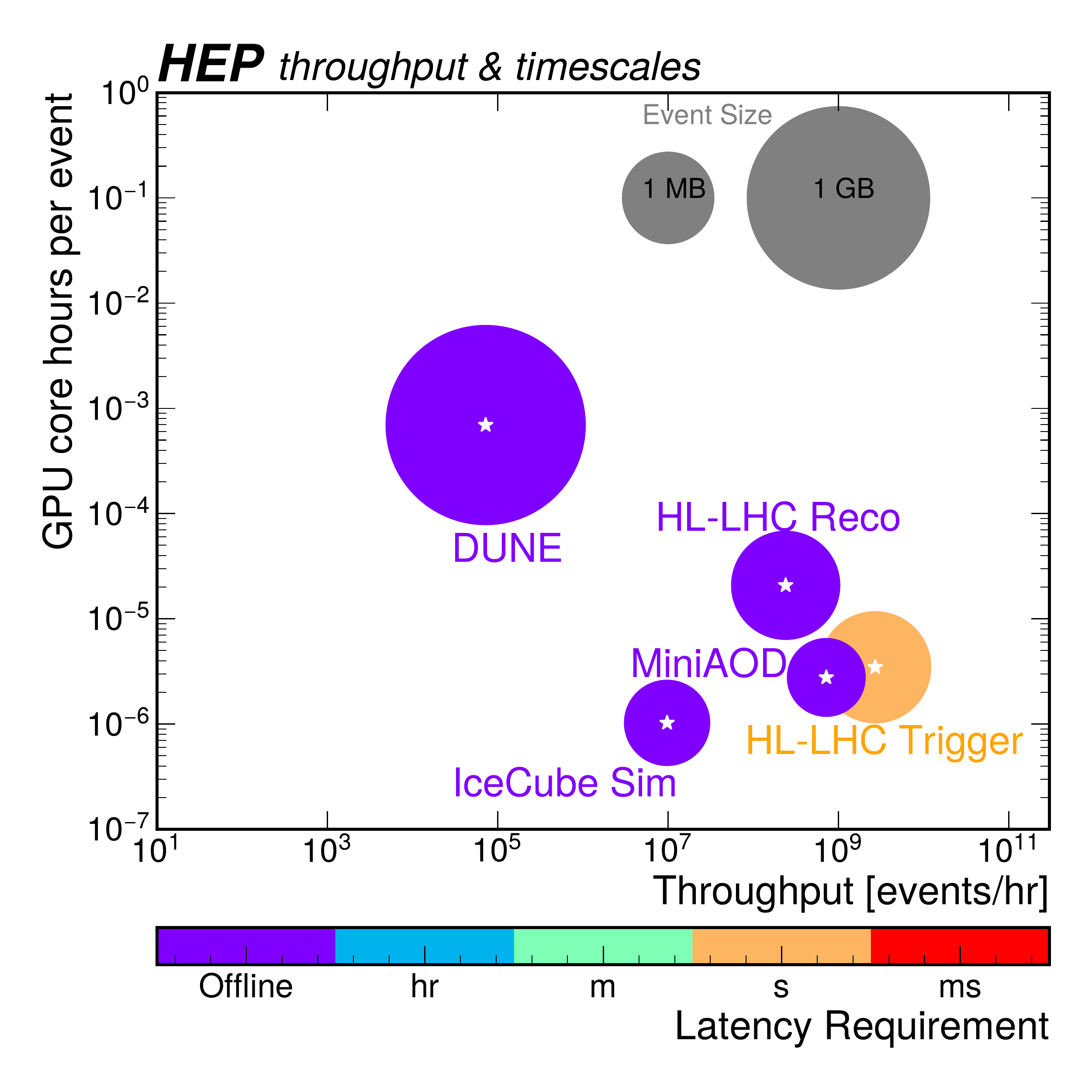}
    \caption{Throughput and CPU (left) and GPU (right) core hours per event for highlighted HEP workflows. The size of the circles represents typical event sizes, and their colors represent latency requirements (per event) for the workflows.}
    \label{fig:hep_times_and_compute}
\end{figure}



Computing demands in high energy physics (HEP) are rapidly growing, as experiments increase data-taking rates and detector complexity.
For example, as the Large Hadron Collider (LHC)~\cite{Evans:2008zzb} transitions to the High-Luminosity LHC (HL-LHC) era, the data-taking rates for the CMS~\cite{CMS:2008xjf} and ATLAS~\cite{ATLAS:2008xda} experiments are expected to increase by a factor of 7--10.
To address concerns that demand for computing resources would outstrip availability, both experiments have launched R\&D efforts to increase computational efficiency of their workflows~\cite{Software:2815292, Collaboration:2802918}.

One pathway to reducing the need for CPUs is to employ coprocessors, such as GPUs and FPGAs, and to create algorithms that efficiently take advantage of these coprocessors' large numbers of processing units and inherently parallelized designs.
In particular, these architectures are becoming increasingly popular for accelerating calculations in ML algorithms.
Within HEP, ML algorithms are widely used for regression and classification tasks, and coprocessors can eliminate the need for CPU resources to perform inference for these algorithms.
Several applications of ML and coprocessor-based acceleration are highlighted in the following paragraphs.

\textbf{Offline refinement of LHC data:} An example workflow used by the CMS experiment that can take advantage of GPU acceleration is MiniAOD production.
MiniAOD production is a data slimming and enhancement step executed for the full dataset typically a few times per year, because various algorithms within the workflow are occasionally updated~\cite{Petrucciani:2015gjw}.
Within this workflow, three main algorithms can be easily ported to run on GPUs:
\begin{itemize}
    \item ParticleNet~\cite{Qu:2019gqs}, which is a graph neural network for jet tagging and regression that represents jets as ``particle clouds''.
    \item DeepMET~\cite{DeepMET}, which is a deep neural network model that estimates the $\ensuremath{{\vec p}_{\mathrm{T}}^{\kern1pt\text{miss}}}$ in an LHC event.
    \item DeepTau~\cite{CMS:2022prd}, which is a deep neural network model to identify hadronically decaying tau leptons from jets.
\end{itemize}
Normally, a full MiniAOD processing takes about 2 days and occupies about 200,000 CPU cores as it runs over the full dataset, which consists of more than 10 billion LHC events.
The algorithms cited above constitute about 10\% of the total per-event latency, which is about 1 second per event.
When these algorithms run on GPUs, they execute about 10 times faster, effectively eliminating this latency from the workflow.
The models discussed here generate about 10-20 GB/s of network traffic when SONIC (Section~\ref{sec:overlaps}) is used with a compute farm of 40,000 CPU threads.

\textbf{Online and offline LHC event reconstruction:} One of the most important aspects of event reconstruction at the LHC is charged particle trajectory reconstruction, or ``tracking''.
When processing events at the LHC, tracking can consume about half of the per-event latency, and this latency increases dramatically with the tracking detector occupancy, as traditional algorithms compare all allowed hit combinations, naively $\mathcal{O}(n^2)$.
A graph neural network called Exa.TrkX~\cite{ExaTrkX} has been introduced to rapidly find correct combinations of hits to create tracks; this algorithm runs about 20 times faster on a GPU and has close to linear scaling.
The ATLAS experiment is exploring the integration of versions of Exa.TrkX into both local reconstruction, which is the first offline step of analysis, and their triggering workflow.
This is especially relevant to tracking in the HL-LHC era, where the number of tracks will increase by roughly an order of magnitude.

In addition to applying ML for tracking, recent efforts have been made to apply ML algorithms to calorimeter clustering. Calorimeters are designed to capture and measure the energy of incident particles, and when a particle interacts with the calorimeter, it typically creates a ``shower'' as it deposits energy, resulting in measurable energy in many sensitive elements of the calorimeters (``cells''). Clustering algorithms are designed to link multiple cells together, such that the sum of the cells' energy measurements approximates the energy deposited by the original particle. Traditionally, these are domain algorithms that perform loops over calorimeter cells, considering adjacency and energy patterns, potentially involving multiple steps. In the HL-LHC era, it is unclear if these traditional algorithms can achieve adequate physics performance while satisfying computing constraints. Because of this, recent efforts have been made to create ML-based clustering algorithms that can take advantage of modern hardware acceleration.  For example, a graph neural network for clustering that also comprises a noise filter has demonstrated good performance for the CMS upgrade High Granularity Calorimeter (HGCAL)~\cite{Bhattacharya_2023}, which has much finer spatial resolution than the current CMS calorimeters. ATLAS has also developed a GPU-based porting of topological clustering for calorimeters, which can execute clustering about 4 times faster than CPU-based clustering and could be deployed in their online high-level trigger~\cite{DosSantosFernandes:2802139}. 

Full data processing is typically performed about once per year, occupying hundreds of thousands of CPU cores for several days.
Per-event latency could be as high as about 3 seconds for HL-LHC events, which could be dramatically reduced by running tracking on GPUs.
Similarly, the CPU-based component of the trigger will have to process over 500,000 events per second.
Executing this component of the trigger takes on the order of 100,000 CPU threads, so each thread should process an event in roughly 200 ms.
The algorithms run in the trigger step are reduced in complexity relative to offline processing in order to improve latency. Adding computational power through the use of GPUs and ML algorithms would improve the overall complexity of the trigger system allowing for algorithms that more closely replicate the offline computing system. 

\textbf{Particle classification at DUNE:} Of course, the use of ML algorithms is not unique to LHC-based experiments.
In the data-processing workflow of the ProtoDUNE-SP experiment, which is a liquid argon time projection chamber prototype of the DUNE far detector, a convolutional neural network (CNN) is used to identify track– and shower–like particles and Michel electrons~\cite{Abed_Abud_2022}.
This CNN takes an image with a typical size of 4 GB as input, and it consumes about two-thirds of the ProtoDUNE reconstruction latency per event when the whole workflow is run on CPUs~\cite{Wang:2020fjr}.
When the CNN is run on GPUs, it is 18 times faster and its latency is reduced to 10\% of the whole workflow~\cite{DuneOnGPU}.
DUNE data processing is an offline reconstruction workflow that is planned to be run for the full dataset once per year. This whole-dataset reprocessing requires several days to complete, with one event taking about 25 seconds to process when heterogeneous computing resources are used.
The workflow takes advantage of remote GPU resources using the SONIC framework discussed in Section~\ref{sec:overlaps}.
Roughly 1000 CPU threads are used in DUNE data processing, which generates about 100 GB/s of network traffic into model hosting servers.

\textbf{Data processing and simulation for IceCube:} IceCube is a kilometer-scale neutrino detector in Antarctica that detects the Cherenkov radiation produced by particle collisions within Antarctic ice sheets~\cite{icecube1}.
The major background to neutrino events in IceCube are cosmic rays, which are about 500,000 times more prevalent, occuring at a rate of $10^{10}$ per year~\cite{iceprod}.
IceCube runs with an event rate of about 3000 Hz, taking in about 1 TB of information per day (at about 20 kB per event) and reducing this to about 100 GB, with much of the data processing compute power dedicated to neutrino vs. cosmic ray discrimination.
This processing occurs on 300 CPU cores located in Antarctica and 100 cores located in Wisconsin.
There is also a data stream dedicated to identifying astrophysical phenomena, such as supernovae~\cite{icecube_supernova1, icecube_supernova2}.
When IceCube detects such events, it can coordinate with other observatories (thus contributing to the multi-messenger astronomy scheme discussed previously), so processing of these events must be completed promptly, typically within seconds.
There are about 100,000 events per year that are processed in this data stream.
Currently, ML algorithms are not used in IceCube's online data processing, though such algorithms are currently being developed.
By running photon propagation algorithms on GPUs, IceCube events can be simulated in 3.7 ms, representing acceleration by a factor of 200 relative to running on CPU alone~\cite{icecube_gpu}.
With these improvements, IceCube can simulate events at a rate that matches the incoming data of about 3000 Hz.

\textbf{Anomaly detection:} A final example where ML can play a major role in HEP is in anomaly detection.
In recent years, many ML-based algorithms have emerged to try to detect Beyond the Standard Model (BSM) physics events~\cite{Metodiev_2017, Collins_2019, Kasieczka_2021, Amram_2021, Hallin_2022, Park_2021, Jawahar_2022}.
Many of these either train a neural network to distinguish events in a signal region from events in a sideband region or attempt to encode Standard Model physics into an autoencoder, such that BSM events would not be well reconstructed from the latent space.
In particular, there are ongoing efforts to deploy autoencoder algorithms on FPGAs~\cite{Govorkova_2022}.
Because FPGAs perform inference so quickly, it is possible that these anomaly detection algorithms could be run at trigger-level for LHC experiments (in the hardware-based trigger component, rather than the CPU-based component discussed earlier), allowing them to comb through a much larger dataset than only events that have already passed the trigger system.






\pagebreak
\section{Software}
\label{sec:overlaps}
A major concern when deploying workflows on heterogeneous architectures is efficient use of resources.
The most straightforward approach for efficient coprocessor usage would be to purchase or reserve machines with the ``correct'' amount of coprocessor resources, such that the coprocessor will not be saturated when the target workflow runs.
Each CPU in the machine will communicate with the coprocessor, and for a given coprocessor and CPU type, there will be some optimal ratio of CPU to coprocessor where the coprocessor is almost, but not completely, saturated.

While this approach is easy to conceptualize, it has a few drawbacks.
First, workflows change as a function of time, so the optimal CPU to coprocessor ratio is likely to change as algorithms evolve.
Thus, if machines are \textit{purchased} with a particular specification, they can quickly become outdated.
Furthermore, those machines would have been optimized for a single workflow, and when that workflow is not running, those machines will either sit idle or be used inefficiently by another workflow.
On the other hand, machines can be reserved through various services, such as Google Cloud or Amazon Web Services.
While these services do provide highly customizable machines, they can incur significant recurring costs depending on how frequently the workflow must be run, as well as data ingress or egress needs.
It would also likely be difficult to use cloud resources for online workflows that require low latency, as data transfer between the detector site and the cloud site could simply take too long or consume too much network bandwidth.

An alternate paradigm, Inference as a Service (IaaS), has recently gained some traction in HEP and GW physics experiments.
In the IaaS scheme, coprocessor resources are factorized out of CPU machines: CPU-based \textit{clients} send inference requests with necessary input and metadata to coprocessor-providing \textit{servers} via network calls.
Algorithm execution is performed on the server, and inference results are sent back to the client again via a network call.
In this way, a coprocessor can communicate with any number of client CPUs, making it highly flexible, as the optimal CPU to coprocessor ratio can be achieved for any workflow, assuming there is a sufficiently large pool of coprocessor resources.
It also has the simple benefit of allowing CPU-only machines to take advantage of coprocessor-based acceleration.

In HEP, an IaaS design pattern called ``Services for Optimized Network Inference on Coprocessors'' (SONIC) has been introduced~\cite{Krupa:2020bwg}, and has already been incorporated into the CMS software framework, CMSSW, and the LArSoft framework used by protoDUNE.
SONIC takes advantage of pre-existing industry efforts, and, for example, uses the NVIDIA Triton Inference Server~\cite{triton} to host models and provide inference.
Depending on the workflow and the experiment software framework capabilities, SONIC can run with asynchronous non-blocking calls or synchronously.
In CMSSW, SONIC can make asynchronous non-blocking calls, and any latency introduced by remote calls has been shown to be negligible for client to server distances of at least 100 miles.
SONIC has also been introduced into the DUNE workflow, but this implementation is synchronous.
Here, the advantages of running on GPU are so significant that latency from call time is unimportant.
It is generally true that latency from remote calls is small, but one can still factor this effect into performance projections.
Both CMS and DUNE have deployed SONIC for large-scale production workflows in the cloud.
In particular, groups from both experiments started server clusters of 100 GPUs each (behind Kubernetes load balancers), and observed expected speed-ups in their workflows.

In GW physics, recent developments have been made for streaming inference on time-series data, with tools like {\tt{hermes}}.\footnote{\url{https://github.com/ML4GW/hermes}} It also adopts the IaaS paradigm, using NVIDIA Triton Inference Server infrastructure with efficient data snapshotting to perform inference on only new time points in an overlapping time window. This has been shown to demonstrate millisecond time inference for data cleaning using DeepClean in an online setting~\cite{gunny2022hardware}. The {\tt{hermes}} infrastructure is also adopted in generic deep learning-based online searches like Aframe and MLy~\cite{2020arXiv200914611S}.

Currently, SONIC and {\tt{hermes}} rely on the use of containers, in particular Singularity or Docker, so it is important that any computing site where workflows will be deployed supports the required software.
Many groups looking to use SONIC also plan on using Kubernetes for automating deployment and scaling, as it naturally integrates with SONIC's containers.
For many fields, support for package management systems, like Conda is useful. Similarly, support for mainstream ML backends, such as TensorFlow, PyTorch, and ONNX are also needed.
Many HEP and GW workflows also rely on CernVM-File System (CVMFS) for software distribution and Globus, Rucio, etc. for data distribution, so it is useful when this is available on worker nodes.
Lastly, a batch job deployment framework, such as Slurm or HTCondor, is also needed for large-scale job deployment.

In the case of EM and GW, communicating discovery alerts and data products between observatories is crucial for the success of multi-messenger astronomy. To this end, streaming tools based on Apache Kafka~\footnote{\url{https://kafka.apache.org/}} have been developed by efforts such as SCiMMA HOPSKOTCH~\footnote{\url{https://scimma.org/}}. HOPSKOTCH is a scalable, high-throughput and low-latency platform for handling real-time data streams for multi-messenger astronomy. While alert brokers can ingest this data, and toolkits such as TOM~\footnote{\url{https://lco.global/tomtoolkit/}} and Treasure Map~\footnote{\url{http://treasuremap.space/}} can help to coordinate follow-up resources, prioritizing follow-up relies on algorithms run on high-performance computers. Efforts such as LINCC~\footnote{\url{https://www.lsstcorporation.org/lincc/}} are helping to develop the necessary software infrastructure for processing the data streams.  Provisioning the necessary software on HPC systems to ingest, perform the inference, publish, and archive results will be crucial in the future of joint follow-up from multiple observatories.

As mentioned in Section~\ref{sec:hep_sec}, there are ongoing efforts to deploy ML algorithms on coprocessors other than GPUs, such as FPGAs. Technologies such as FPGAs and application-specific integrated circuits (ASICs) can provide high inference speeds in an energy-efficient manner, but it can be more difficult to implement algorithms on these platforms. To simplify deployment, the {\tt{hls4ml}}\footnote{\url{https://github.com/fastmachinelearning/hls4ml}} (``high level synthesis for machine learning'') framework has been introduced, which provides many tools to make algorithms compatible with hardware constraints~\cite{fahim2021hls4ml}. FPGAs have been used via an IaaS scheme through the FPGAs-as-a-Service Toolkit (FaaST)~\cite{Rankin_2020}, demonstrating the dramatic acceleration of a small neural net for calorimeter energy regression and a much larger ResNet-50 algorithm. In this case, {\tt{hls4ml}} was used to write the FPGA kernels.

\section{Computing}
\label{sec:apps}
In order to achieve workflow acceleration via heterogeneous computing, it is necessary to have access to appropriate coprocessor resources.
While some large-scale experiments have sufficient budgets to make large-scale coprocessor purchases, this is not the case for all experiments.
Additionally, if workflows that use coprocessors are not run frequently enough, it may not be justified to acquire coprocessors in the first place. R\&D for large-scale heterogeneous workflows is typically performed \textit{before} any purchase and often requires access to large-scale resources to test scaling behavior.

Two possibilities for ephemeral large-scale coprocessor access are the cloud and high-performance computing centers (HPC centers).
Cloud resources are generally highly configurable, and with some effort, virtual machines can be configured to run most software and have personalized batch submission clusters.
However, this requires expertise and time from researchers, which is not always desirable.

At HPC centers, there is usually less configurability, as external researchers are not granted root access and there are networking firewalls because of justifiable security concerns.
HPC centers also typically have their own batch submission systems and only support certain container software.

If a future physics-oriented computing cluster were to be created, the following outline would meet the needs of the algorithms highlighted in this white paper.
\begin{itemize}
    \item Compute scale: A cluster with roughly 300,000 CPU threads would be able to service offline and online data processing needs of many experiments. This would match current distributed computing core availability for a large-scale experiment, such as CMS or ATLAS~\cite{Balcas:2297171}, and should provide adequate resources as long as experiments are able to stagger large-scale processing campaigns.
    \item Heterogeneous compute power: About 1,000 GPUs would be needed to meet the needs of these workflows. As a current scale reference, the CMS experiment recently acquired a high-level trigger (HLT) farm consisting of 200 dual processor servers, each equipped with two AMD EPYC “Milan” 7763 CPUs and two NVIDIA Tesla T4 GPUs, thus totaling 400 GPUs~\cite{Bocci_2023}. This farm is designed to handle online processing of the HLT for LHC Run 3. It has a ratio of hyperthreaded CPU cores to GPUs of 128:1. The proposed physics-oriented cluster should be able to provide adequate coprocessor resources for many large and small scale experiments, and the proposed scale should be sufficient for the current online processing projects of experiments discussed in Sections~\ref{sec:em_sec}--\ref{sec:hep_sec}, with some resources left over for other offline work. As workflows evolve in the future, it is possible that the number of GPUs at the cluster will need to increase. Similarly, large-scale data processing that occurs approximately annually for many experiments may need to be staggered and scheduled during downtimes for other experiments, when online demands are lower.
    \item Flexibility for future architectures: The IaaS paradigm allows for the use of coprocessors other than GPUs, such as FPGAs or IPUs. As architectures are developed to accelerate particular algorithm classes, it would be beneficial if the cluster retains the capability to add resources with new and unique architectures.
    \item Node-to-node networking: An internal network capable of handling at least 200 GB/s is required to enable inference as a service at large scale. The higher the bandwidth, the more workflows could be executed simultaneously.
    \item External networking: For online workflows, experimenters must stream data into and out of the computing site, making this a critical consideration for a physics computing cluster.
    \item Software support: The software requirements for the communities included in this white paper are addressed at the end of Section~\ref{sec:overlaps}.
    \item Data availability: While this has already been somewhat achieved in HEP and GW, the analysis workflows are not immediately portable from one computing center to another. Developing new tools or improving upon existing tools to enable this portability will be necessary for future large-scale experimental physics.
\end{itemize}


\pagebreak
\section{Outlook}
\label{sec:outlook}

\begin{figure}[h]
    \centering
    \includegraphics[width=1.0\textwidth]{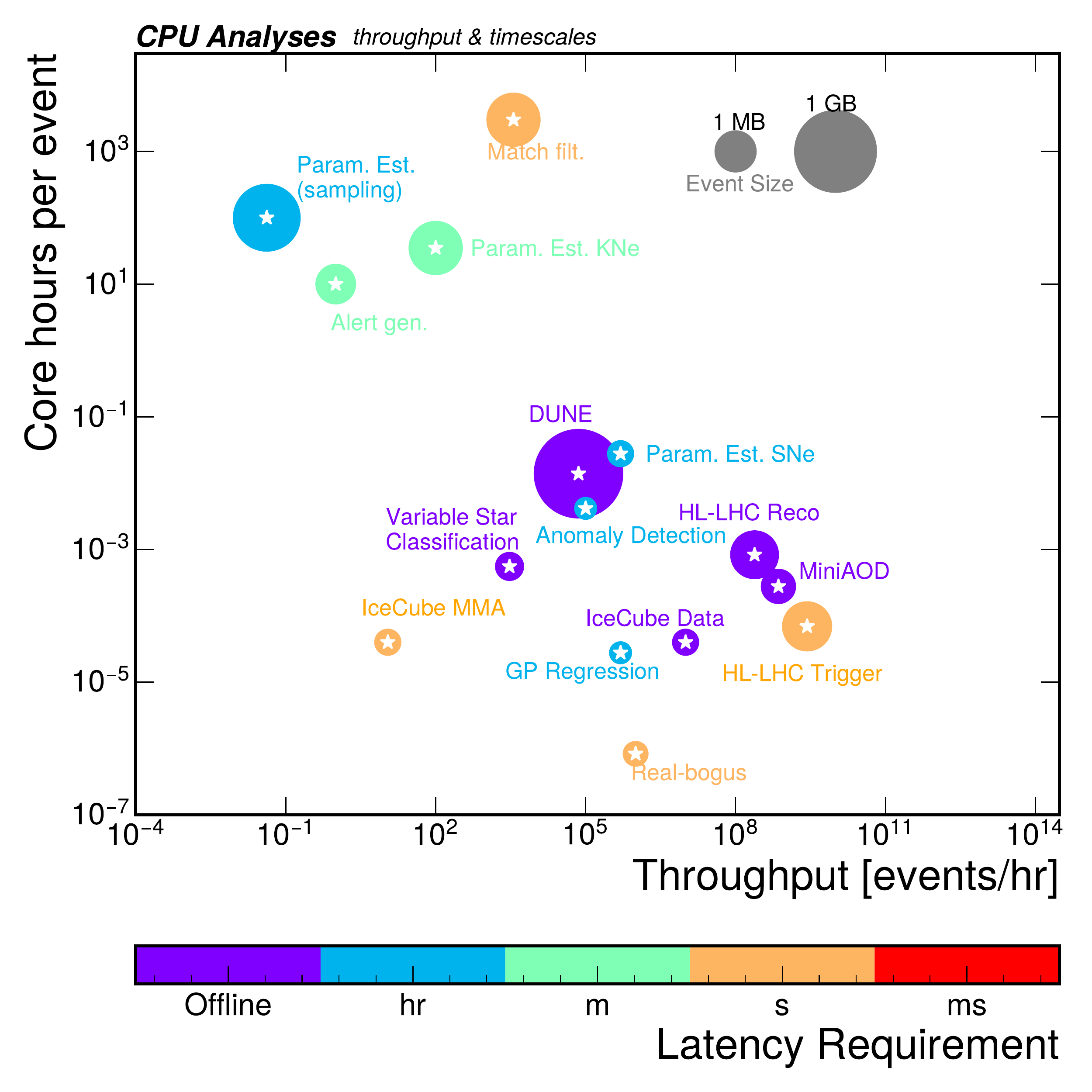}
    \caption{Throughput and CPU core hours per event for highlighted workflows across disciplines. The size of the circles represents typical event sizes, and their colors represent latency requirements (per event) for the workflows.}
    \label{fig:combined_analyses_cpu}
\end{figure}

\begin{figure}[h]
    \centering
    \includegraphics[width=1.0\textwidth]{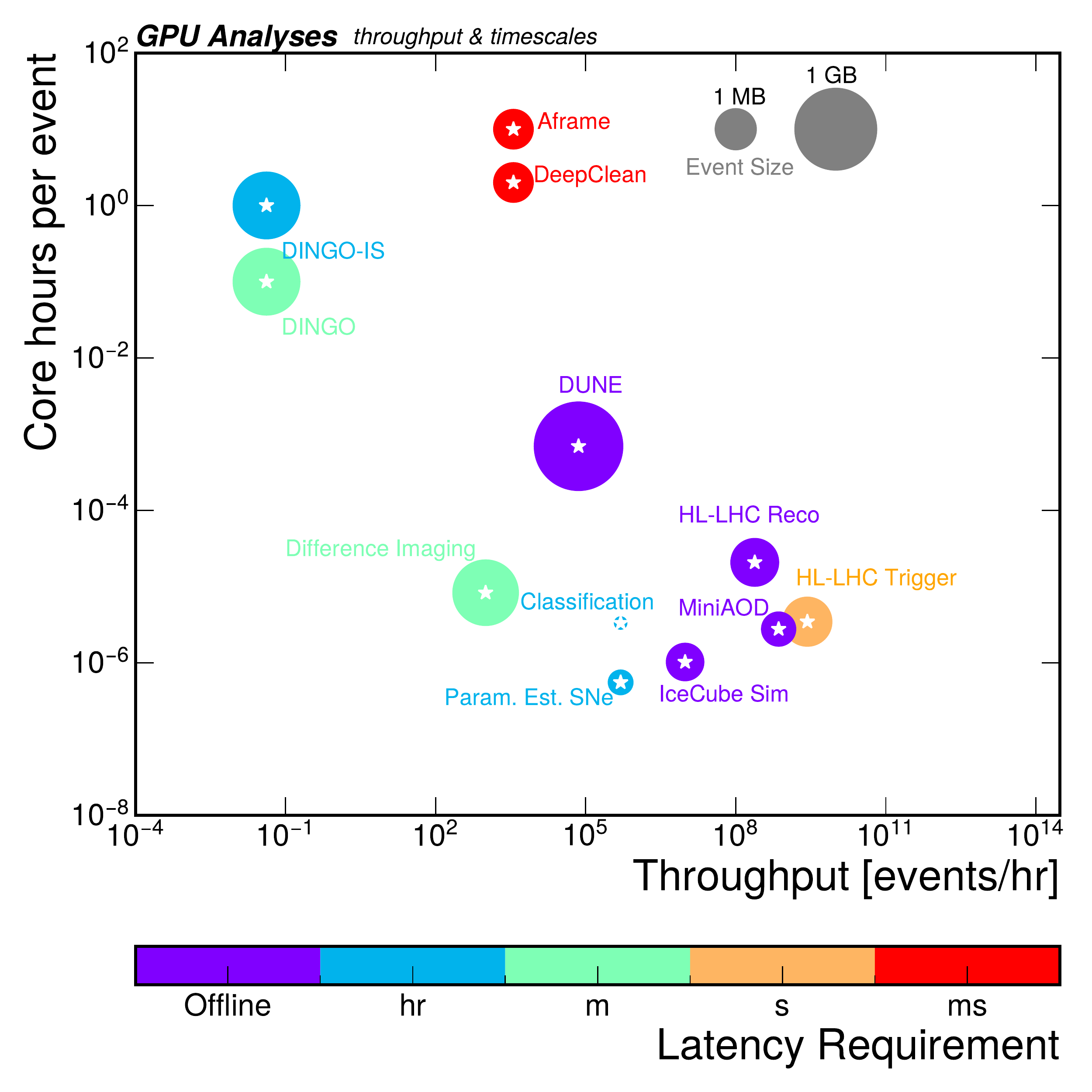}
    \caption{Throughput and GPU core hours per event for highlighted workflows across disciplines. The size of the circles represents typical event sizes, and their colors represent latency requirements (per event) for the workflows.}
    \label{fig:combined_analyses_gpu}
\end{figure}

The incorporation of ML and AI algorithms into workflows and the use of heterogeneous computing are increasingly common features in modern experimental physics, especially as collaborations strive for greater computing efficiency.
Across and within disciplines, there is a wide diversity of computing needs, spanning many orders of magnitude in core requirements, latency requirements, bandwidth, and volume.
This diversity is illustrated in Figs.~\ref{fig:combined_analyses_cpu}~and~\ref{fig:combined_analyses_gpu}.
A computing site with sufficient hardware capabilities \textit{and} appropriate software libraries that can meet the needs of the different experimental communities highlighted in this white paper would serve to benefit this community and the wider scientific community.
With no single computing site capable of satisfying the current needs of all the experiments outlined in this whitepaper, individual experiments have been deploying their own specialized computing clusters, incurring significant financial and labor costs.
If a large-scale, physics-dedicated HPC center were to be established in the future, it would facilitate cross-disciplinary synergies, enable rapid workflow research and development, and provide resources for cutting-edge experiments conducting large-volume data processing.
Ultimately, we believe that such a development would bring substantial benefits to the physics community as a whole.

\pagebreak
\section*{Acknowledgments}
We acknowledge the NSF HD Planning grant for support of the conference that led to this work.

\clearpage
\bibliographystyle{frontiersFPHY} 
\bibliography{references}

\begin{thebibliography}{81}
\expandafter\ifx\csname natexlab\endcsname\relax\def\natexlab#1{#1}\fi
\expandafter\ifx\csname urlstyle\endcsname\relax
  \expandafter\ifx\csname doi\endcsname\relax
  \def\doi#1{doi:\discretionary{}{}{}#1}\fi \else
  \expandafter\ifx\csname doi\endcsname\relax
  \def\doi{doi:\discretionary{}{}{}\begingroup \urlstyle{rm}\Url}\fi \fi
\expandafter\ifx\csname selectlanguage\endcsname\relax
  \def\selectlanguage#1{}\fi

\bibitem[{{Pang} et~al.(2022){Pang}, {Dietrich}, {Coughlin}, {Bulla}, {Tews},
  {Almualla} et~al.}]{Pang2022KNe}
{Pang} PTH, {Dietrich} T, {Coughlin} MW, {Bulla} M, {Tews} I, {Almualla} M,
  et~al.
\newblock {NMMA: A nuclear-physics and multi-messenger astrophysics framework
  to analyze binary neutron star mergers}.
\newblock {\em arXiv e-prints\/}  (2022) arXiv:2205.08513.
\newblock \doi{10.48550/arXiv.2205.08513}.

\bibitem[{{Boone}(2021)}]{Boone2021Parsnip}
{Boone} K.
\newblock {ParSNIP: Generative Models of Transient Light Curves with
  Physics-enabled Deep Learning}.
\newblock {\em \aj\/} {\bf 162} (2021) 275.
\newblock \doi{10.3847/1538-3881/ac2a2d}.

\bibitem[{{Villar}(2022)}]{Villar2022SBI}
{Villar} VA.
\newblock {Amortized Bayesian Inference for Supernovae in the Era of the Vera
  Rubin Observatory Using Normalizing Flows}.
\newblock {\em arXiv e-prints\/}  (2022) arXiv:2211.04480.
\newblock \doi{10.48550/arXiv.2211.04480}.

\bibitem[{{Muthukrishna} et~al.(2022){Muthukrishna}, {Mandel}, {Lochner},
  {Webb}, and {Narayan}}]{Muthukrishna2022}
{Muthukrishna} D, {Mandel} KS, {Lochner} M, {Webb} S, {Narayan} G.
\newblock {Real-time detection of anomalies in large-scale transient surveys}.
\newblock {\em \mnras\/} {\bf 517} (2022) 393--419.
\newblock \doi{10.1093/mnras/stac2582}.

\bibitem[{{Audenaert} et~al.(2021){Audenaert}, {Kuszlewicz}, {Handberg},
  {Tkachenko}, {Armstrong}, {Hon} et~al.}]{Audenaert2021}
{Audenaert} J, {Kuszlewicz} JS, {Handberg} R, {Tkachenko} A, {Armstrong} DJ,
  {Hon} M, et~al.
\newblock {TESS Data for Asteroseismology (T'DA) Stellar Variability
  Classification Pipeline: Setup and Application to the Kepler Q9 Data}.
\newblock {\em \aj\/} {\bf 162} (2021) 209.
\newblock \doi{10.3847/1538-3881/ac166a}.

\bibitem[{{Boone}(2019)}]{Boone2019Avocado}
{Boone} K.
\newblock {Avocado: Photometric Classification of Astronomical Transients with
  Gaussian Process Augmentation}.
\newblock {\em \aj\/} {\bf 158} (2019) 257.
\newblock \doi{10.3847/1538-3881/ab5182}.

\bibitem[{{Muthukrishna} et~al.(2019){Muthukrishna}, {Narayan}, {Mandel},
  {Biswas}, and {Hlo{\v{z}}ek}}]{Muthukrishna19RAPID}
{Muthukrishna} D, {Narayan} G, {Mandel} KS, {Biswas} R, {Hlo{\v{z}}ek} R.
\newblock {RAPID: Early Classification of Explosive Transients Using Deep
  Learning}.
\newblock {\em \pasp\/} {\bf 131} (2019) 118002.
\newblock \doi{10.1088/1538-3873/ab1609}.

\bibitem[{{Duev} et~al.(2019){Duev}, {Mahabal}, {Masci}, {Graham}, {Rusholme},
  {Walters} et~al.}]{Duev2019Real-bogus}
{Duev} DA, {Mahabal} A, {Masci} FJ, {Graham} MJ, {Rusholme} B, {Walters} R,
  et~al.
\newblock {Real-bogus classification for the Zwicky Transient Facility using
  deep learning}.
\newblock {\em \mnras\/} {\bf 489} (2019) 3582--3590.
\newblock \doi{10.1093/mnras/stz2357}.

\bibitem[{{Corbett} et~al.(2022){Corbett}, {Vasquez Soto}, {Machia},
  {Galliher}, {Gonzalez}, and {Law}}]{Corbett2022DIA}
{Corbett} H, {Vasquez Soto} A, {Machia} L, {Galliher} N, {Gonzalez} R, {Law}
  NM.
\newblock {The sky at one terabit per second: architecture and implementation
  of the Argus Array Hierarchical Data Processing System}.
\newblock {\em Society of Photo-Optical Instrumentation Engineers (SPIE)
  Conference Series\/} (2022), vol. 12189, 1218910.
\newblock \doi{10.1117/12.2629533}.

\bibitem[{{Ivezi{\'c}} et~al.(2019){Ivezi{\'c}}, {Kahn}, {Tyson}, {Abel},
  {Acosta}, {Allsman} et~al.}]{Ivezic2009LSST:Products}
{Ivezi{\'c}} {\v Z}, {Kahn} SM, {Tyson} JA, {Abel} B, {Acosta} E, {Allsman} R,
  et~al.
\newblock {LSST: From Science Drivers to Reference Design and Anticipated Data
  Products}.
\newblock {\em \apj\/} {\bf 873} (2019) 111.
\newblock \doi{10.3847/1538-4357/ab042c}.

\bibitem[{{Bellm} et~al.(2019){Bellm}, {Kulkarni}, {Graham}, {Dekany}, {Smith},
  {Riddle} et~al.}]{Bellm2019ZTF}
{Bellm} EC, {Kulkarni} SR, {Graham} MJ, {Dekany} R, {Smith} RM, {Riddle} R,
  et~al.
\newblock {The Zwicky Transient Facility: System Overview, Performance, and
  First Results}.
\newblock {\em \pasp\/} {\bf 131} (2019) 018002.
\newblock \doi{10.1088/1538-3873/aaecbe}.

\bibitem[{{Ricker} et~al.(2015){Ricker}, {Winn}, {Vanderspek}, {Latham},
  {Bakos}, {Bean} et~al.}]{TESS_Ricker_2015}
{Ricker} GR, {Winn} JN, {Vanderspek} R, {Latham} DW, {Bakos} G{\'A}, {Bean} JL,
  et~al.
\newblock {Transiting Exoplanet Survey Satellite (TESS)}.
\newblock {\em Journal of Astronomical Telescopes, Instruments, and Systems\/}
  {\bf 1} (2015) 014003.
\newblock \doi{10.1117/1.JATIS.1.1.014003}.

\bibitem[{{Chambers} et~al.(2016){Chambers}, {Magnier}, {Metcalfe},
  {Flewelling}, {Huber}, {Waters} et~al.}]{Chambers2016Panstarrs}
{Chambers} KC, {Magnier} EA, {Metcalfe} N, {Flewelling} HA, {Huber} ME,
  {Waters} CZ, et~al.
\newblock {The Pan-STARRS1 Surveys}.
\newblock {\em arXiv e-prints\/}  (2016) arXiv:1612.05560.

\bibitem[{{F{\"o}rster} et~al.(2021){F{\"o}rster}, {Cabrera-Vives},
  {Castillo-Navarrete}, {Est{\'e}vez}, {S{\'a}nchez-S{\'a}ez}, {Arredondo}
  et~al.}]{ALeRCE_Broker}
{F{\"o}rster} F, {Cabrera-Vives} G, {Castillo-Navarrete} E, {Est{\'e}vez} PA,
  {S{\'a}nchez-S{\'a}ez} P, {Arredondo} J, et~al.
\newblock {The Automatic Learning for the Rapid Classification of Events
  (ALeRCE) Alert Broker}.
\newblock {\em \aj\/} {\bf 161} (2021) 242.
\newblock \doi{10.3847/1538-3881/abe9bc}.

\bibitem[{{Nordin} et~al.(2019){Nordin}, {Brinnel}, {van Santen}, {Bulla},
  {Feindt}, {Franckowiak} et~al.}]{AMPEL_Broker}
{Nordin} J, {Brinnel} V, {van Santen} J, {Bulla} M, {Feindt} U, {Franckowiak}
  A, et~al.
\newblock {Transient processing and analysis using AMPEL: alert management,
  photometry, and evaluation of light curves}.
\newblock {\em \aap\/} {\bf 631} (2019) A147.
\newblock \doi{10.1051/0004-6361/201935634}.

\bibitem[{{Narayan} et~al.(2018){Narayan}, {Zaidi}, {Soraisam}, {Wang},
  {Lochner}, {Matheson} et~al.}]{ANTARES_Broker}
{Narayan} G, {Zaidi} T, {Soraisam} MD, {Wang} Z, {Lochner} M, {Matheson} T,
  et~al.
\newblock {Machine-learning-based Brokers for Real-time Classification of the
  LSST Alert Stream}.
\newblock {\em \apjs\/} {\bf 236} (2018) 9.
\newblock \doi{10.3847/1538-4365/aab781}.

\bibitem[{{M{\"o}ller} et~al.(2021){M{\"o}ller}, {Peloton}, {Ishida},
  {Arnault}, {Bachelet}, {Blaineau} et~al.}]{Fink_Broker}
{M{\"o}ller} A, {Peloton} J, {Ishida} EEO, {Arnault} C, {Bachelet} E,
  {Blaineau} T, et~al.
\newblock {FINK, a new generation of broker for the LSST community}.
\newblock {\em \mnras\/} {\bf 501} (2021) 3272--3288.
\newblock \doi{10.1093/mnras/staa3602}.

\bibitem[{{Smith} et~al.(2019){Smith}, {Williams}, {Young}, {Ibsen}, {Smartt},
  {Lawrence} et~al.}]{LASAIR_Broker}
{Smith} KW, {Williams} RD, {Young} DR, {Ibsen} A, {Smartt} SJ, {Lawrence} A,
  et~al.
\newblock {Lasair: The Transient Alert Broker for LSST:UK}.
\newblock {\em Research Notes of the American Astronomical Society\/} {\bf 3}
  (2019) 26.
\newblock \doi{10.3847/2515-5172/ab020f}.

\bibitem[{{Villar} et~al.(2020){Villar}, {Hosseinzadeh}, {Berger}, {Ntampaka},
  {Jones}, {Challis} et~al.}]{Villar2020SuperRAENN}
{Villar} VA, {Hosseinzadeh} G, {Berger} E, {Ntampaka} M, {Jones} DO, {Challis}
  P, et~al.
\newblock {SuperRAENN: A Semisupervised Supernova Photometric Classification
  Pipeline Trained on Pan-STARRS1 Medium-Deep Survey Supernovae}.
\newblock {\em \apj\/} {\bf 905} (2020) 94.
\newblock \doi{10.3847/1538-4357/abc6fd}.

\bibitem[{{Villar} et~al.(2021){Villar}, {Cranmer}, {Berger}, {Contardo}, {Ho},
  {Hosseinzadeh} et~al.}]{Villar2021_Anomalydetection}
{Villar} VA, {Cranmer} M, {Berger} E, {Contardo} G, {Ho} S, {Hosseinzadeh} G,
  et~al.
\newblock {A Deep-learning Approach for Live Anomaly Detection of Extragalactic
  Transients}.
\newblock {\em \apjs\/} {\bf 255} (2021) 24.
\newblock \doi{10.3847/1538-4365/ac0893}.

\bibitem[{{M{\"o}ller} and {de Boissi{\`e}re}(2020)}]{SupernnoovaMoller2019}
{M{\"o}ller} A, {de Boissi{\`e}re} T.
\newblock {SuperNNova: an open-source framework for Bayesian, neural
  network-based supernova classification}.
\newblock {\em \mnras\/} {\bf 491} (2020) 4277--4293.
\newblock \doi{10.1093/mnras/stz3312}.

\bibitem[{{Pimentel} et~al.(2023){Pimentel}, {Est{\'e}vez}, and
  {F{\"o}rster}}]{Pimentel2023}
{Pimentel} {\'O}, {Est{\'e}vez} PA, {F{\"o}rster} F.
\newblock {Deep Attention-based Supernovae Classification of Multiband Light
  Curves}.
\newblock {\em \aj\/} {\bf 165} (2023) 18.
\newblock \doi{10.3847/1538-3881/ac9ab4}.

\bibitem[{{Allam} et~al.(2023){Allam}, {Peloton}, and {McEwen}}]{Tarek2023}
{Allam} J Tarek, {Peloton} J, {McEwen} JD.
\newblock {The Tiny Time-series Transformer: Low-latency High-throughput
  Classification of Astronomical Transients using Deep Model Compression}.
\newblock {\em arXiv e-prints\/}  (2023) arXiv:2303.08951.
\newblock \doi{10.48550/arXiv.2303.08951}.

\bibitem[{{Mandel} et~al.(2022){Mandel}, {Thorp}, {Narayan}, {Friedman}, and
  {Avelino}}]{Mandel2022}
{Mandel} KS, {Thorp} S, {Narayan} G, {Friedman} AS, {Avelino} A.
\newblock {A hierarchical Bayesian SED model for Type Ia supernovae in the
  optical to near-infrared}.
\newblock {\em \mnras\/} {\bf 510} (2022) 3939--3966.
\newblock \doi{10.1093/mnras/stab3496}.

\bibitem[{Abbott et~al.(2016)Abbott, Abbott, Abbott, Abernathy, Acernese,
  Ackley et~al.}]{Abbott_2016}
Abbott B, Abbott R, Abbott T, Abernathy M, Acernese F, Ackley K, et~al.
\newblock Observation of gravitational waves from a binary black hole merger.
\newblock {\em Physical Review Letters\/} {\bf 116} (2016).
\newblock \doi{10.1103/physrevlett.116.061102}.

\bibitem[{{Einstein}(1916)}]{1916AnP...354..769E}
{Einstein} A.
\newblock {Die Grundlage der allgemeinen Relativit{\"a}tstheorie}.
\newblock {\em Annalen der Physik\/} {\bf 354} (1916) 769--822.
\newblock \doi{10.1002/andp.19163540702}.

\bibitem[{Abbott et~al.(2021)}]{LIGOScientific:2021djp}
Abbott R, et~al.
\newblock {GWTC-3: Compact Binary Coalescences Observed by LIGO and Virgo
  During the Second Part of the Third Observing Run}  (2021).

\bibitem[{{National Academies of Sciences, Engineering, and
  Medicine}(2021)}]{astro2020}
{National Academies of Sciences, Engineering, and Medicine}.
\newblock {\em Pathways to Discovery in Astronomy and Astrophysics for the
  2020s\/} (Washington, DC: The National Academies Press) (2021).
\newblock \doi{10.17226/26141}.

\bibitem[{Abbott et~al.(2017)Abbott, Abbott, Abbott, Acernese, Ackley, Adams
  et~al.}]{PhysRevLett.119.161101}
Abbott B, Abbott R, Abbott T, Acernese F, Ackley K, Adams C, et~al.
\newblock {GW170817}: Observation of gravitational waves from a binary neutron
  star inspiral.
\newblock {\em Physical Review Letters\/} {\bf 119} (2017).
\newblock \doi{10.1103/physrevlett.119.161101}.

\bibitem[{Evans et~al.(2021)Evans, Adhikari, Afle, Ballmer, Biscoveanu,
  Borhanian et~al.}]{evans2021horizon}
Evans M, Adhikari RX, Afle C, Ballmer SW, Biscoveanu S, Borhanian S, et~al.
\newblock A horizon study for cosmic explorer: Science, observatories, and
  community  (2021).

\bibitem[{Amaro-Seoane et~al.(2017)Amaro-Seoane, Audley, Babak, Baker,
  Barausse, Bender et~al.}]{amaroseoane2017laser}
Amaro-Seoane P, Audley H, Babak S, Baker J, Barausse E, Bender P, et~al.
\newblock Laser interferometer space antenna  (2017).

\bibitem[{Ormiston et~al.(2020)Ormiston, Nguyen, Coughlin, Adhikari, and
  Katsavounidis}]{Ormiston_2020}
Ormiston R, Nguyen T, Coughlin M, Adhikari RX, Katsavounidis E.
\newblock Noise reduction in gravitational-wave data via deep learning.
\newblock {\em Physical Review Research\/} {\bf 2} (2020).
\newblock \doi{10.1103/physrevresearch.2.033066}.

\bibitem[{Beveridge~D(2023)}]{beveridge}
Beveridge~D WA Wen~L.
\newblock Detection of {B}inary {B}lack {H}ole {M}ergers from the
  {S}ignal-to-{N}oise {R}atio {T}ime {S}eries {U}sing deep learning. (in
  preparation)  (2023).

\bibitem[{Chatterjee et~al.(2021)Chatterjee, Wen, Diakogiannis, and
  Vinsen}]{Chatterjee_2021}
Chatterjee C, Wen L, Diakogiannis F, Vinsen K.
\newblock Extraction of binary black hole gravitational wave signals from
  detector data using deep learning.
\newblock {\em Physical Review D\/} {\bf 104} (2021).
\newblock \doi{10.1103/physrevd.104.064046}.

\bibitem[{Chatterjee and Wen(2022)}]{chatterjee2022premerger}
Chatterjee C, Wen L.
\newblock Pre-merger sky localization of gravitational waves from binary
  neutron star mergers using deep learning  (2022).

\bibitem[{Chatterjee et~al.(2022)Chatterjee, Wen, Beveridge, Diakogiannis, and
  Vinsen}]{chatterjee2022rapid}
Chatterjee C, Wen L, Beveridge D, Diakogiannis F, Vinsen K.
\newblock Rapid localization of gravitational wave sources from compact binary
  coalescences using deep learning  (2022).

\bibitem[{Guo et~al.(2022)Guo, Williams, Heng, Gabbard, Bae, Kang
  et~al.}]{Guo_2022}
Guo W, Williams D, Heng IS, Gabbard H, Bae YB, Kang G, et~al.
\newblock Mimicking mergers: mistaking black hole captures as mergers.
\newblock {\em Monthly Notices of the Royal Astronomical Society\/} {\bf 516}
  (2022) 3847--3860.
\newblock \doi{10.1093/mnras/stac2385}.

\bibitem[{Alsing et~al.(2019)Alsing, Charnock, Feeney, and
  Wandelt}]{Alsing_2019}
Alsing J, Charnock T, Feeney S, Wandelt B.
\newblock Fast likelihood-free cosmology with neural density estimators and
  active learning.
\newblock {\em Monthly Notices of the Royal Astronomical Society\/}  (2019).
\newblock \doi{10.1093/mnras/stz1960}.

\bibitem[{{Zhang} et~al.(2021){Zhang}, {Bloom}, {Gaudi}, {Lanusse}, {Lam}, and
  {Lu}}]{2021AJ....161..262Z}
{Zhang} K, {Bloom} JS, {Gaudi} BS, {Lanusse} F, {Lam} C, {Lu} JR.
\newblock {Real-time Likelihood-free Inference of Roman Binary Microlensing
  Events with Amortized Neural Posterior Estimation}.
\newblock {\em Astronomical Journal\/} {\bf 161} (2021) 262.
\newblock \doi{10.3847/1538-3881/abf42e}.

\bibitem[{Cranmer et~al.(2020)Cranmer, Brehmer, and Louppe}]{Cranmer_2020}
Cranmer K, Brehmer J, Louppe G.
\newblock The frontier of simulation-based inference.
\newblock {\em Proceedings of the National Academy of Sciences\/} {\bf 117}
  (2020) 30055--30062.
\newblock \doi{10.1073/pnas.1912789117}.

\bibitem[{Dax et~al.(2021)Dax, Green, Gair, Macke, Buonanno, and
  Sch\"olkopf}]{PhysRevLett.127.241103}
Dax M, Green SR, Gair J, Macke JH, Buonanno A, Sch\"olkopf B.
\newblock Real-time gravitational wave science with neural posterior
  estimation.
\newblock {\em Phys. Rev. Lett.\/} {\bf 127} (2021) 241103.
\newblock \doi{10.1103/PhysRevLett.127.241103}.

\bibitem[{Dax et~al.(2022)Dax, Green, Gair, Deistler, Sch\"olkopf, and
  Macke}]{Dax:2021myb}
Dax M, Green SR, Gair J, Deistler M, Sch\"olkopf B, Macke JH.
\newblock {Group equivariant neural posterior estimation}.
\newblock {\em International Conference on Learning Representations\/} (2022).

\bibitem[{Dax et~al.(2023{\natexlab{a}})Dax, Green, Gair, P\"urrer, Wildberger,
  Macke et~al.}]{PhysRevLett.130.171403}
Dax M, Green SR, Gair J, P\"urrer M, Wildberger J, Macke JH, et~al.
\newblock Neural importance sampling for rapid and reliable gravitational-wave
  inference.
\newblock {\em Phys. Rev. Lett.\/} {\bf 130} (2023{\natexlab{a}}) 171403.
\newblock \doi{10.1103/PhysRevLett.130.171403}.

\bibitem[{Dax et~al.(2023{\natexlab{b}})Dax, Wildberger, Buchholz, Green,
  Macke, and Schölkopf}]{dax2023flow}
Dax M, Wildberger J, Buchholz S, Green SR, Macke JH, Schölkopf B.
\newblock Flow matching for scalable simulation-based inference
  (2023{\natexlab{b}}).

\bibitem[{McLeod et~al.(2022)McLeod, Jacobs, Chatterjee, Wen, and
  Panther}]{mcleod2022rapid}
McLeod A, Jacobs D, Chatterjee C, Wen L, Panther F.
\newblock Rapid mass parameter estimation of binary black hole coalescences
  using deep learning  (2022).

\bibitem[{Evans and Bryant(2008)}]{Evans:2008zzb}
Evans L, Bryant P.
\newblock {LHC Machine}.
\newblock {\em JINST\/} {\bf 3} (2008) S08001.
\newblock \doi{10.1088/1748-0221/3/08/S08001}.

\bibitem[{Chatrchyan et~al.(2008)}]{CMS:2008xjf}
Chatrchyan S, et~al.
\newblock {The CMS Experiment at the CERN LHC}.
\newblock {\em JINST\/} {\bf 3} (2008) S08004.
\newblock \doi{10.1088/1748-0221/3/08/S08004}.

\bibitem[{Aad et~al.(2008)}]{ATLAS:2008xda}
Aad G, et~al.
\newblock {The ATLAS Experiment at the CERN Large Hadron Collider}.
\newblock {\em JINST\/} {\bf 3} (2008) S08003.
\newblock \doi{10.1088/1748-0221/3/08/S08003}.

\bibitem[{{CMS Offline Software and Computing}(2022)}]{Software:2815292}
{CMS Offline Software and Computing}.
\newblock {CMS Phase-2 Computing Model: Update Document}.
\newblock Tech. rep., CERN, Geneva (2022).

\bibitem[{{ATLAS Collaboration}(2022)}]{Collaboration:2802918}
{ATLAS Collaboration}.
\newblock {ATLAS Software and Computing HL-LHC Roadmap}.
\newblock Tech. rep., CERN, Geneva (2022).

\bibitem[{Petrucciani et~al.(2015)Petrucciani, Rizzi, and
  Vuosalo}]{Petrucciani:2015gjw}
Petrucciani G, Rizzi A, Vuosalo C.
\newblock {Mini-AOD: A New Analysis Data Format for CMS}.
\newblock {\em J. Phys. Conf. Ser.\/} {\bf 664} (2015) 7.
\newblock \doi{10.1088/1742-6596/664/7/072052}.

\bibitem[{Qu and Gouskos(2020)}]{Qu:2019gqs}
Qu H, Gouskos L.
\newblock {ParticleNet: Jet Tagging via Particle Clouds}.
\newblock {\em Phys. Rev. D\/} {\bf 101} (2020) 056019.
\newblock \doi{10.1103/PhysRevD.101.056019}.

\bibitem[{Feng(2020)}]{DeepMET}
Feng Y.
\newblock {\em A New Deep-Neural-Network--Based Missing Transverse Momentum
  Estimator, and its Application to W Recoil\/}.
\newblock Ph.D. thesis, University of Maryland, College Park (2020).
\newblock \doi{10.13016/e6ze-zycc}.

\bibitem[{{CMS Collaboration}(2022)}]{CMS:2022prd}
{CMS Collaboration}.
\newblock {Identification of hadronic tau lepton decays using a deep neural
  network}.
\newblock {\em JINST\/} {\bf 17} (2022) P07023.
\newblock \doi{10.1088/1748-0221/17/07/P07023}.

\bibitem[{Lazar et~al.(2022)Lazar, Ju, Murnane, Calafiura, Farrell, Xu
  et~al.}]{ExaTrkX}
Lazar A, Ju X, Murnane D, Calafiura P, Farrell S, Xu Y, et~al.
\newblock Accelerating the inference of the {Exa.TrkX} pipeline  (2022).
\newblock \doi{10.48550/ARXIV.2202.06929}.

\bibitem[{Bhattacharya et~al.(2023)Bhattacharya, Chernyavskaya, Ghosh, Gray,
  Kieseler, Klijnsma et~al.}]{Bhattacharya_2023}
Bhattacharya S, Chernyavskaya N, Ghosh S, Gray L, Kieseler J, Klijnsma T,
  et~al.
\newblock {GNN}-based end-to-end reconstruction in the {CMS} phase 2
  high-granularity calorimeter.
\newblock {\em Journal of Physics: Conference Series\/} {\bf 2438} (2023)
  012090.
\newblock \doi{10.1088/1742-6596/2438/1/012090}.

\bibitem[{Dos Santos~Fernandes(2022)}]{DosSantosFernandes:2802139}
Dos Santos~Fernandes N.
\newblock {GPU acceleration of the ATLAS calorimeter clustering algorithm}.
\newblock Tech. rep., CERN, Geneva (2022).

\bibitem[{Abud et~al.(2022)}]{Abed_Abud_2022}
Abud AA, et~al.
\newblock Separation of track- and shower-like energy deposits in
  {ProtoDUNE}-{SP} using a convolutional neural network.
\newblock {\em The European Physical Journal C\/} {\bf 82} (2022).
\newblock \doi{10.1140/epjc/s10052-022-10791-2}.

\bibitem[{Wang et~al.(2021)Wang, Yang, Acosta~Flechas, Harris, Hawks, Holzman
  et~al.}]{Wang:2020fjr}
Wang M, Yang T, Acosta~Flechas M, Harris P, Hawks B, Holzman B, et~al.
\newblock {GPU-Accelerated Machine Learning Inference as a Service for
  Computing in Neutrino Experiments}.
\newblock {\em Front. Big Data\/} {\bf 3} (2021) 604083.
\newblock \doi{10.3389/fdata.2020.604083}.

\bibitem[{Cai et~al.(2023)Cai, Herner, Yang, Wang, Flechas, Harris
  et~al.}]{DuneOnGPU}
Cai T, Herner K, Yang T, Wang M, Flechas MA, Harris P, et~al.
\newblock {Accelerating Machine Learning Inference with GPUs in ProtoDUNE Data
  Processing}  (2023).
\newblock \doi{10.48550/ARXIV.2301.04633}.

\bibitem[{Halzen and Klein(2010)}]{icecube1}
Halzen F, Klein SR.
\newblock Invited review article: {IceCube}: An instrument for neutrino
  astronomy.
\newblock {\em Review of Scientific Instruments\/} {\bf 81} (2010) 081101.
\newblock \doi{10.1063/1.3480478}.

\bibitem[{Aartsen et~al.(2015{\natexlab{a}})}]{iceprod}
Aartsen M, et~al.
\newblock The {IceProd} framework: Distributed data processing for the
  {IceCube} neutrino observatory.
\newblock {\em Journal of Parallel and Distributed Computing\/} {\bf 75}
  (2015{\natexlab{a}}) 198--211.
\newblock \doi{10.1016/j.jpdc.2014.08.001}.

\bibitem[{Köpke et~al.(2018)}]{icecube_supernova1}
Köpke L, et~al.
\newblock Improved detection of supernovae with the {IceCube} observatory.
\newblock {\em Journal of Physics: Conference Series\/} {\bf 1029} (2018)
  012001.
\newblock \doi{10.1088/1742-6596/1029/1/012001}.

\bibitem[{Aartsen et~al.(2015{\natexlab{b}})}]{icecube_supernova2}
Aartsen MG, et~al.
\newblock {THE} {DETECTION} {OF} a {SN} {IIn} {IN} {OPTICAL} {FOLLOW}-{UP}
  {OBSERVATIONS} {OF} {ICECUBE} {NEUTRINO} {EVENTS}.
\newblock {\em The Astrophysical Journal\/} {\bf 811} (2015{\natexlab{b}}) 52.
\newblock \doi{10.1088/0004-637x/811/1/52}.

\bibitem[{Schwanekamp et~al.(2022)Schwanekamp, Hohl, Chirkin
  et~al.}]{icecube_gpu}
Schwanekamp H, Hohl R, Chirkin D, et~al.
\newblock {Accelerating IceCube’s Photon Propagation Code with CUDA}.
\newblock {\em Comput. Softw. Big Sci.\/} {\bf 6} (2022) 1.
\newblock \doi{10.1007/s41781-022-00080-8}.

\bibitem[{Metodiev et~al.(2017)Metodiev, Nachman, and Thaler}]{Metodiev_2017}
Metodiev EM, Nachman B, Thaler J.
\newblock Classification without labels: learning from mixed samples in high
  energy physics.
\newblock {\em Journal of High Energy Physics\/} {\bf 2017} (2017).
\newblock \doi{10.1007/jhep10(2017)174}.

\bibitem[{Collins et~al.(2019)Collins, Howe, and Nachman}]{Collins_2019}
Collins JH, Howe K, Nachman B.
\newblock Extending the search for new resonances with machine learning.
\newblock {\em Physical Review D\/} {\bf 99} (2019).
\newblock \doi{10.1103/physrevd.99.014038}.

\bibitem[{Kasieczka et~al.(2021)Kasieczka, Nachman, Shih, Amram, Andreassen,
  Benkendorfer et~al.}]{Kasieczka_2021}
Kasieczka G, Nachman B, Shih D, Amram O, Andreassen A, Benkendorfer K, et~al.
\newblock The {LHC} olympics 2020 a community challenge for anomaly detection
  in high energy physics.
\newblock {\em Reports on Progress in Physics\/} {\bf 84} (2021) 124201.
\newblock \doi{10.1088/1361-6633/ac36b9}.

\bibitem[{Amram and Suarez(2021)}]{Amram_2021}
Amram O, Suarez CM.
\newblock {Tag N' Train}: a technique to train improved classifiers on
  unlabeled data.
\newblock {\em Journal of High Energy Physics\/} {\bf 2021} (2021).
\newblock \doi{10.1007/jhep01(2021)153}.

\bibitem[{Hallin et~al.(2022)Hallin, Isaacson, Kasieczka, Krause, Nachman,
  Quadfasel et~al.}]{Hallin_2022}
Hallin A, Isaacson J, Kasieczka G, Krause C, Nachman B, Quadfasel T, et~al.
\newblock Classifying anomalies through outer density estimation.
\newblock {\em Physical Review D\/} {\bf 106} (2022).
\newblock \doi{10.1103/physrevd.106.055006}.

\bibitem[{Park et~al.(2021)Park, Rankin, Udrescu, Yunus, and
  Harris}]{Park_2021}
Park SE, Rankin D, Udrescu SM, Yunus M, Harris P.
\newblock Quasi anomalous knowledge: searching for new physics with embedded
  knowledge.
\newblock {\em Journal of High Energy Physics\/} {\bf 2021} (2021).
\newblock \doi{10.1007/jhep06(2021)030}.

\bibitem[{Jawahar et~al.(2022)Jawahar, Aarrestad, Chernyavskaya, Pierini,
  Wozniak, Ngadiuba et~al.}]{Jawahar_2022}
Jawahar P, Aarrestad T, Chernyavskaya N, Pierini M, Wozniak KA, Ngadiuba J,
  et~al.
\newblock Improving variational autoencoders for new physics detection at the
  {LHC} with normalizing flows.
\newblock {\em Frontiers in Big Data\/} {\bf 5} (2022).
\newblock \doi{10.3389/fdata.2022.803685}.

\bibitem[{Govorkova et~al.(2022)Govorkova, Puljak, Aarrestad, James, Loncar,
  Pierini et~al.}]{Govorkova_2022}
Govorkova E, Puljak E, Aarrestad T, James T, Loncar V, Pierini M, et~al.
\newblock Autoencoders on field-programmable gate arrays for real-time,
  unsupervised new physics detection at 40 {MHz} at the {Large Hadron
  Collider}.
\newblock {\em Nature Machine Intelligence\/} {\bf 4} (2022) 154--161.
\newblock \doi{10.1038/s42256-022-00441-3}.

\bibitem[{Krupa et~al.(2021)}]{Krupa:2020bwg}
Krupa J, et~al.
\newblock {GPU coprocessors as a service for deep learning inference in high
  energy physics}.
\newblock {\em Mach. Learn. Sci. Tech.\/} {\bf 2} (2021) 035005.
\newblock \doi{10.1088/2632-2153/abec21}.

\bibitem[{NVIDIA(2018)}]{triton}
NVIDIA.
\newblock {NVIDIA Triton Inference Server}  (2018).
\newblock Accessed: 2022-09-07.

\bibitem[{Gunny et~al.(2022)Gunny, Rankin, Krupa, Saleem, Nguyen, Coughlin
  et~al.}]{gunny2022hardware}
Gunny A, Rankin D, Krupa J, Saleem M, Nguyen T, Coughlin M, et~al.
\newblock Hardware-accelerated inference for real-time gravitational-wave
  astronomy.
\newblock {\em Nature Astronomy\/} {\bf 6} (2022) 529--536.

\bibitem[{{Skliris} et~al.(2020){Skliris}, {Norman}, and
  {Sutton}}]{2020arXiv200914611S}
{Skliris} V, {Norman} MRK, {Sutton} PJ.
\newblock {Real-Time Detection of Unmodelled Gravitational-Wave Transients
  Using Convolutional Neural Networks}.
\newblock {\em arXiv e-prints\/}  (2020) arXiv:2009.14611.
\newblock \doi{10.48550/arXiv.2009.14611}.

\bibitem[{Fahim et~al.(2021)Fahim, Hawks, Herwig, Hirschauer, Jindariani, Tran
  et~al.}]{fahim2021hls4ml}
Fahim F, Hawks B, Herwig C, Hirschauer J, Jindariani S, Tran N, et~al.
\newblock hls4ml: An open-source codesign workflow to empower scientific
  low-power machine learning devices  (2021).

\bibitem[{Rankin et~al.(2020)Rankin, Krupa, Harris, Flechas, Holzman, Klijnsma
  et~al.}]{Rankin_2020}
Rankin D, Krupa J, Harris P, Flechas MA, Holzman B, Klijnsma T, et~al.
\newblock {{FPGAs}-as-a-Service Toolkit ({FaaST})}.
\newblock {\em 2020 {IEEE}/{ACM} International Workshop on Heterogeneous
  High-performance Reconfigurable Computing (H2RC)\/} ({IEEE}) (2020).
\newblock \doi{10.1109/h2rc51942.2020.00010}.

\bibitem[{Balcas et~al.(2017)Balcas, Bockelman, Hufnagel, Hurtado~Anampa,
  Aftab~Khan, Larson et~al.}]{Balcas:2297171}
Balcas J, Bockelman B, Hufnagel D, Hurtado~Anampa K, Aftab~Khan F, Larson K,
  et~al.
\newblock {Stability and scalability of the CMS Global Pool: Pushing HTCondor
  and glideinWMS to new limits}.
\newblock {\em J. Phys.: Conf. Ser.\/} {\bf 898} (2017) 052031.
\newblock \doi{10.1088/1742-6596/898/5/052031}.

\bibitem[{Bocci and on~behalf of~the CMS~Collaboration(2023)}]{Bocci_2023}
Bocci A, on~behalf of~the CMS~Collaboration.
\newblock {CMS} high level trigger performance comparison on {CPUs} and {GPUs}.
\newblock {\em Journal of Physics: Conference Series\/} {\bf 2438} (2023)
  012016.
\newblock \doi{10.1088/1742-6596/2438/1/012016}.

\end{thebibliography}
\end{document}